\documentclass[12pt]{article}
\usepackage{amssymb}
\usepackage{amsmath}
\usepackage{eucal}
\usepackage{epsfig}

\def\be{\begin{equation}}
\def\eeq{\end{equation}}
\def\bea{\begin{eqnarray}}
\def\eea{\end{eqnarray}}

\def\re#1{(\ref{#1})}

\def\abar{\bar{\alpha}}
\def\kbar{\bar{\kappa}}
\def\kabar{\bar{\kappa}}
\def\kbar{\bar{k}}
\def\lbar{\bar{\lambda}}
\def\mbar{\bar{\mu}}
\def\nbar{\bar{\nu}}
\def\pbar{\bar{p}}
\def\sbar{\bar{s}}
\def\Rbar{\bar{R}}
\def\rbar{\bar{\rho}}
\def\zbar{\bar{\zeta}}

\begin{document}

\begin{titlepage}
\rightline{hep-th/yymmnnn}
\vskip 2. cm
\begin{center}
\huge{\bf{Warped Kaluza-Klein Towers Revisited}}
\end{center}
\vskip 0.6 cm
\begin{center} {\Large{Fernand Grard$^1$}},
\ \ {\Large{Jean Nuyts$^2$}}
\end{center}
\vskip 1 cm

\noindent{\bf Abstract}
\vskip 0.2 cm
{\small
\noindent
Inspired by the warped Randall Sundrum scenario proposed to solve
the mass scale hierarchy problem
with a compactified fifth extra dimension, a similar model
with no metric singularities has been elaborated.
In this framework, the  Kaluza-Klein reduction equations
for a real massless scalar field propagating in the bulk
have been studied
carefully from the point of view of hermiticity so as to formulate in a mathematically
rigorous way all the possible boundary conditions and corresponding mass eigenvalue
towers and tachyon states.
The physical masses as observable in our four-dimensional brane are deduced
from these mass eigenvalues depending on the location of the brane on the extra dimension
axis.
Examples of mass towers and tachyons and related field probability densities
are presented from numerical computations performed for some arbitrary choices of the
parameters of the model.
 }
\vfill
\noindent
{\it $^1$  Fernand.Grard@umh.ac.uk,  Physique G\'en\'erale et
Physique des Particules El\'ementaires,
Universit\'e de Mons-Hainaut, 20 Place du Parc, 7000 Mons, Belgium}
\vskip 0.2 cm
\noindent
{\it $^2$  Jean.Nuyts@umh.ac.be,
Physique Th\'eorique et Math\'ematique,
Universit\'e de Mons-Hainaut,
20 Place du Parc, 7000 Mons, Belgium}
\end{titlepage}

\section{Introduction}{\label{Intro}}

In a previous paper \cite{GN}, we reanalysed mathematically,
within the Arkhani-Ahmed, Dimopoulos, Dvali \cite{ADD}
large extra dimensions
model, the procedure of generation of the Kaluza-Klein masses \cite{KK},
stressing that it is the momentum squared
in the extra dimensions (and not the momentum itself) which is the
physically relevant quantity and hence corresponds to an operator
which must essentially be hermitian.
For illustration purpose, we restricted ourselves
to the case of a five-dimensional
massless
real scalar field supposed
to propagate in the flat five-dimensional bulk.
The extra dimension
is compactified to a finite range, say $[0,2\pi R]$,
either on a circle (then $R$ is
interpreted
as the radius of the circle) or on a finite strip
(then $2\pi R=L$ is the length of the strip).
All the allowed boundary conditions resulting
from the requirement that the
extra dimension momentum squared must be
a  mathematically precisely-defined
symmetric operator have been established (see also
\cite{AGM}).
We deduced from them, besides
the usual regularly spaced Kaluza-Klein mass towers,
new towers with non regular mass spacing and tachyons.
These considerations
should be extended to vector and tensor fields.

In this article, inspired by the Randall and Sundrum scenario \cite{RS},
we have developed a model based on a warped space with one extra dimension
and basic parameters chosen so as to solve similarly the mass scale hierarchy problem.

In Randall-Sundrum, the fifth dimension $s$ is compactified
to an orbifold of radius $R$. A so called
Planck brane is located at $s=0$ 
while the TeV brane or Standard Model brane is at
$s=\pi R$. We depart from the original scenario,
postulating that the compactification is on a strip,
that the metric has no singularity and that only one particular brane,
the TeV brane on which we live as a four dimensional observer,
has to be considered physically.

In this framework and again restricting to a real 
massless scalar field propagating in the bulk
we have carefully studied the hermiticity
properties of the operators in the Kaluza-Klein reduction equations
for the adopted five dimensional warped space. We have enumerated
all the allowed boundary conditions and from them
we have deduced the corresponding Kaluza-Klein mass eigenstate towers and
tachyon states and have studied their main properties.

As will be shown, the values of the observable
physical masses in the Kaluza-Klein
towers can be deduced from the mass eigenvalues and depend
on the particular location of our four dimensional brane in $s$
as do the field probability densities which
are related (when dynamics and kinematics are included) to the
overall probability that the associated mass states
would appear to the observer.

In the five dimensional bulk, we also postulate that
all the dimensionfull parameters
are scaled with a unique mass, the Planck mass $M_{Pl}$.
It then happens that,
by an adequate choice of the reduced parameters
defining the model,
all the low lying physical masses
obtained from the eigenvalues of
the Kaluza-Klein reduction equation are of order TeV for
an observer living in our 4-dimensional brane.

\section{Operators in the five-dimensional warped space.
Mathematical considerations}

The warped five dimensional space
with coordinates $x^A, A=0,1,2,3,5$
is composed of a flat $SO(1,3)$ invariant
infinite four-dimensional subspace
labeled by $x^{\mu}$ ($\mu=0,1,2,3$)
with signature $(+,-,-,-)$ and a
spacelike fifth dimension with
coordinate $x^5\equiv s$
on the finite strip $0 \leq s\leq 2\pi R$. The metric, unique up to rescaling,
\begin{equation}
dS^2=g_{AB}\,dx^A dx^B=e^{-2ks} dx_{\mu}dx^{\mu}-ds^2
\label{metric}
\end{equation}
satisfies Einstein's equations with
a stress-energy tensor identically zero and
a bulk negative cosmological constant $\Lambda$
as the unique origin of the induced Riemann metric \re{metric}.
Indeed $k$, here chosen positive
(see the discussion about the sign of $k$ in Appendix\re{appendixB}),
and $\Lambda$ are related by
\begin{equation}
k=\sqrt{-\frac{\Lambda}{6}}\ ,\quad k>0\ .
\label{klambda}
\end{equation}

A free massless scalar field $\Phi(x,s)$ in this warped space satisfies the
invariant equation
\begin{equation}
\square_{\rm{Riemann}}\Phi\equiv
\frac{1}{\sqrt{g}}\partial_A \sqrt{g}g^{AB}\partial_B\Phi =0\ .
\label{dalemb5}
\end{equation}
From the metric \re{metric}, $\sqrt{g}=e^{-4ks}$ and Eq.\re{dalemb5}
becomes
\begin{equation}
\biggl(e^{2ks} \square_4
- e^{4ks}\partial_s e^{-4ks}\partial_s\biggr)\Phi(x^{\mu},s) =0
\label{warpbasic}
\end{equation}
where $\square_4=\partial_{\mu}\partial^{\mu}$
is the usual four dimensional d'Alembertian operator .

We now carry on
a careful study of the hermiticity properties
of the operators appearing in \re{warpbasic}.
For scalar fields, the invariant scalar product
is
\begin{equation}
\bigl(\Psi,\Phi\bigr)
=\int_{-\infty}^{+\infty} d^4x \int_0^{2\pi R}
ds\ \sqrt{g}\   \Psi^{*}(x,s)\,\Phi(x,s) \ .
\label{scalprod}
\end{equation}

Remember that an operator $A$
with dense domain $D(A)$

\begin{itemize}

\item
is symmetric for a scalar product
if
\begin{equation}
\bigl(\Psi,A\Phi\bigr)
=\bigl(A\Psi,\Phi\bigr)
\label{symmetric}
\end{equation}
for all the vectors $\Psi\in D(A)$ and $\Phi\in D(A)$,
i.e. if the adjoint operator $A^{\dagger}$
of the operator $A$ is an extension of $A$:
$A^{\dagger}\Phi=A\Phi$ for all $\Phi\in D(A)$
and $D(A^{\dagger})\supset D(A)$,

\item
is self-adjoint if $A^{\dagger}\Phi=A\Phi$ for all $\Phi\in D(A)$ and
moreover $D(A^{\dagger})=D(A)$,
i.e. if the operator is symmetric and if the equation \re{symmetric}
cannot be extended naturally to vectors $\Psi$ outside $D(A)$.

\end{itemize}

We will call a differential operator which is symmetric
up to boundary conditions
a formally symmetric operator.

\vskip 0.5 cm
\noindent{\bf{Symmetric Operators}}

\noindent For the scalar product \re{scalprod}, the operator $\square_{\rm{Riemann}}$
\re{dalemb5}
is formally symmetric.
The two operators
which appear
in \re{warpbasic} have the following properties
\bea
       &A_1\equiv e^{2ks} \square_4
       &\quad {\rm{is\ self\!\!-\!\!adjoint}}
      \nonumber\\
       &A_2\equiv e^{4ks}\partial_s e^{-4ks}\partial_s
       &\quad {\rm{is\ formally\ symmetric}}\ .
\label{ops}
\eea
By partial integration, one finds that
the domain condition
for the formally symmetric operator $A_2$ to be symmetric is
\begin{equation}
\Biggl[ \Bigl(\Psi^*(\partial_s \Phi)
-(\partial_s \Psi^*) \Phi\Bigl)\Biggr]_{2\pi R}
=e^{8\pi kR }\Biggl[\Bigl(\Psi^*(\partial_s \Phi)
-(\partial_s \Psi^*) \Phi\Bigl)\Biggr]_0
\label{BCfinalwarp2}
\end{equation}
which means that any field $\Phi$ and its derivative,
both evaluated at $s=0$ and $s=2\pi R$, have to satisfy the same
specific linear relations. These relations
which express the boundary conditions will be studied carefully later.

\vskip 0.5 cm
\noindent{\bf{Commuting operators}}

\noindent The two operators defined above \re{ops} do not commute
and hence cannot be diagonalized together.

Multiplying equation \re{warpbasic} on the left by $e^{-2ks}$
leads to the following equivalent equation
\be
\biggl(\square_4 - e^{2ks}
\partial_s e^{-4ks}\partial_s\biggr)\Phi(x^{\mu},s) =0
\label{warpbasic2}
\end{equation}
which defines two commuting operators with the following properties
\bea
&B_1 \equiv \square_4
   &\quad {\rm{is\ self\!\!-\!\!adjoint}}
      \nonumber\\
&B_2 \equiv e^{2ks}\partial_s e^{-4ks}\partial_s
    &\quad {\rm{is\ not\ even\ formally\ symmetric}}\ .
\label{ops2}
\eea

\vskip 0.5 cm
\noindent{\bf{Puzzle}}

\noindent So we are facing a puzzle
\begin{itemize}
\item
either the two operators $A_1,A_2$ \re{ops}
are formally symmetric but do not commute,
\item
or the two operators $B_1,B_2$ \re{ops2} commute
but the second is not formally symmetric.
\end{itemize}

\noindent{\bf{Solving the puzzle}}

\noindent The puzzle can be solved remembering
recent discussions about non-hermitian operators having real
eigenvalues \cite{Ben}.
It was shown in \cite{FN}
that in many cases these non hermitian operators
are in fact equivalent to hermitian operators
by a non-unitary change of basis.
This is the case here.
Indeed after some algebra, considering the non-unitary transformation
induced by $V(s)=e^{ks}$, one finds
\bea
{\widetilde{B}}_1&=& V B_1 V^{-1} =B_1
        \nonumber\\
{\widetilde{B}}_2&=&V B_2 V^{-1}
        \nonumber\\
{\widetilde{\Phi}}&=&V\Phi
\label{warpchange}
\eea
with the result that
${\widetilde{B}}_1$ and ${\widetilde{B}}_2$ are
at least formally symmetric operators
for the induced scalar product
\begin{equation}
\bigl({\widetilde{\Psi}},{\widetilde{\Phi}}\bigr)
=\int_{-\infty}^{+\infty} d^4x \int_0^{2\pi R}
ds\ e^{-6ks}\   {\widetilde{\Psi}}^{*}(x,s)\,{\widetilde{\Phi}}(x,s) \ .
\label{scalprod2}
\end{equation}
It is also easy to see that the natural conditions
for the operator ${\widetilde{B_2}}$ to be
symmetric when acting
on the space of vectors ${\widetilde{\Phi}}$ are
\be
\Biggl[\Bigl({\widetilde{\Psi}}^*(\partial_s {\widetilde{\Phi}})
-(\partial_s {\widetilde{\Psi}}^*)
{\widetilde{\Phi}}\Bigl)\Biggr]_{2\pi R}
=e^{12\pi k R}\Biggl[\Bigl({\widetilde{\Psi}}^*
(\partial_s {\widetilde{\Phi}})
-(\partial_s {\widetilde{\Psi}}^*)
{\widetilde{\Phi}}\Bigl)\Biggr]_0\ .
\label{BCtf}
\end{equation}
These conditions turn out to be fully compatible
with those \re{BCfinalwarp2}
obtained for $\Phi(x^{\mu},s)$
from the requirement that $A_2$ is symmetric.

Thus even though the operator $B_2$ is not symmetric,
it is equivalent through a change of basis to a symmetric operator
${\widetilde{B}}_2$ and
hence will produce real eigenvalues 
which will be related
to the Kaluza-Klein tower masses
as will appear below.

\section{The Kaluza-Klein reduction equations}

To simplify the discussion, we concentrate on the particular case of
a real massless scalar field.
The general procedure to solve the basic equation \re{warpbasic2}
along the Kaluza-Klein reduction method is well-known.
One supposes that the field
$\Phi(x^{\mu},s)$ is a linear combination of terms
where the variables $x^{\mu}$ and $s$ separate
\be
\Phi(x^{\mu},s)=\sum_{n}\phi^{[x]}_n(x^{\mu})\,\phi^{[s]}_n(s)\ .
\label{KKreduc}
\end{equation}
Then $\Phi(x^{\mu},s)$ is a solution of \re{warpbasic2} if
\bea
B_1\ \phi^{[x]}_n(x^{\mu})
                  \equiv& \square_4\  \phi^{[x]}_n(x^{\mu})&
                  =-m_n^2\ \phi^{[x]}_n(x^{\mu})
     \label{boxKK}\\
B_2\ \phi^{[s]}_n(s)
               \equiv& e^{2ks}\partial_s e^{-4ks}\partial_s\
               \phi^{[s]}_n(s)&
               =-m_n^2\ \phi^{[s]}_n(s)\ .
     \label{warpKK}
\eea
From the arguments given above one then concludes

\begin{enumerate}

\item
The operators
$B_1$ and $B_2$ commute and can indeed be diagonalized simultaneously

\item
The operator $B_2$ is equivalent to a formally symmetric operator
through a non-unitary change of basis
and hence, taking into account
boundary conditions compatible with \re{BCfinalwarp2},
Eq.\re{warpKK} gets real $m_n^2$ eigenvalues which can be positive,
zero or negative.

\item
The operator $B_1$ is self-adjoint.
By \re{boxKK}, the solutions for $m^2_n>0$
correspond to four-dimensional physical particles, those with
$m^2_n=0$ to four-dimensional massless particles
and those with $m^2_n<0$
to four-dimensional tachyons.

\end{enumerate}

The boundary restrictions \re{BCfinalwarp2} can be conveniently rewritten
\bea
&&\Biggl[ \Bigl(\psi^{[s]}_p(\partial_s \phi^{[s]}_n)
-(\partial_s \psi^{[s]}_p) \phi^{[s]}_n\Bigl)\Biggr]_{2\pi R}
     \nonumber\\
&&\hskip 1 cm=
\Biggl[\Bigl((e^{4\pi kR }\psi^{[s]}_p)
(\partial_s e^{4\pi kR }\phi^{[s]}_n)
-(\partial_s e^{4\pi kR }\psi^{[s]}_p)
(e^{4\pi kR }\phi^{[s]}_n)\Bigl)\Biggr]_0\ .
\label{BCfinalfinal}
\eea
When $k=0$, these restrictions are identical
to the restrictions applicable in the fully flat case which we studied in
\cite{GN}.
As a consequence, the boundary conditions compatible with \re{BCfinalfinal}
can be obtained by simply replacing
$\phi^{[s]}_n(0)$ and $\partial_s \phi^{[s]}(0)$
by respectively $e^{4\pi kR }\phi^{[s]}_n(0)$ and $e^{4\pi kR }
\partial_s \phi^{[s]}(0)$
in the boundary conditions which we listed for the flat case.

In fact, each field must satisfy
the same set of boundary conditions
which consist of at least two linear relations.
This set defines
specific domains in the Hilbert space.
In Table\re{tableBCdeux}, we give all the possible independent sets
of boundary conditions expressed by just two linear relations.
The box condition appears in Case A6.

Boundary conditions expressed with more than two linear relations
can be considered as restrictions applied
to the cases with two relations.
The domain of the operator is then reduced. For three linear relations,
the sets are given in Table\re{tableBCtrois}.
The set involving four relations
consists of
$\phi^{[s]}_n(0)=0,\ \phi^{[s]}_n(2\pi R)=0,
\ \partial_s \phi^{[s]}_n(0)=0,\ \partial_s \phi^{[s]}_n(2\pi R)=0$,
which can only be satisfied by the trivial field
$\phi_n^{[s]}(s)=0$ and can thus be forgotten.

The boundary conditions have to be imposed to the general solutions
of the equation \re{warpKK} which are

\begin{itemize}

\item

For $m_n^2>0$, the solutions
are linear superpositions of the Bessel
functions $J_2$ and $Y_2$ (see Appendix \re{appendixA})
\be
\phi^{[s]}_n(s)=e^{2ks}\Biggl(\sigma_{n} J_2\left(\frac{m_n e^{ks}}{k}\right)
                              +\tau_{n} Y _2\left(\frac{m_n e^{ks}}{k}\right)\Biggr)\ .
     \label{solwarpKK}
\end{equation}

\item

The solution for $m^2_0=0$ is
\be
\phi^{[s]}_0(s)=\sigma_{0} e^{4ks} + \tau_{0}\ .
     \label{solwarpKK0}
\end{equation}

\item

The tachyon solutions for $m_t^2=-h_t^2<0$ are linear
superpositions of the modified
Bessel functions
$I_2$ and $K_2$ (see Appendix \re{appendixA})
\be
\phi^{[s]}_t(s)=e^{2ks}\Biggl(\sigma_t I_2
                \left(\frac{h_t e^{ks}}{k}\right)
               +\tau_t K _2\left(\frac{h_t e^{ks}}{k}\right)
                                         \Biggr)\ .
     \label{solwarpKKtach}
\end{equation}

\item In the above formulae, $\sigma_{n},\tau_{n},\sigma_{0},\tau_{0},\sigma_{t}$ and
$\tau_{t}$
are constants.

\end{itemize}

\section{Physical considerations. Masses}

We now extend our considerations to potentially
physical consequences including for example
the possible discovery
of TeV warped states at high energy colliders.
In the next subsections, we first discuss
the magnitude of the parameters $k$ and $R$
which occur in the model and also the magnitude of the parameters
$\alpha_i$ ... which define the boundary conditions,
postulating that
there is only one scale in the model, the Planck mass.
We then write explicitly the equations
determining the mass eigenvalues. 
We discuss the interpretation
of these mass eigenvalues in terms of the
physical masses
as they would be observed in our brane.
In particular the
conditions for the existence of zero mass states and of
tachyons mass states are deduced. 
Examples are finally given and discussed.

\subsection{The parameters}

The general philosophy underlying the
warped approach, which was proposed to solve the hierarchy problem,
is
that there is only one mass scale, the Planck mass $M_{Pl}\approx 1.22\ 10^{16}$ TeV,
and hence that any
dimensionfull parameter $p$
with energy dimension $d$ is of order $M_{Pl}^d$. More precisely
\be
p=\pbar\, (M_{Pl})^d
\label{barpara}
\end{equation}
with $\pbar$ a pure number of order one.
In particular, $k=\kbar M_{Pl}$ and $R=\Rbar
M_{Pl}^{-1}$.

As stated above \re{klambda}, the parameter
$k$ is chosen positive.
As will be seen hereafter,
it appears that for an observer
sitting at $s=0$ a reasonable choice for the product $kR$ is
\begin{equation}
kR=\kbar\Rbar\ \approx \ 6.3
\label{kR2}
\end{equation}
as
the resulting lowest masses in the Kaluza-Klein towers
would then be of order 1 TeV.

\subsection{The solutions}

We restrict ourselves to all the boundary conditions
which are expressed by two relations (see Table\re{tableBCdeux}).
We first discuss the
Kaluza-Klein mass towers in general,
then in particular the towers which
have a mass zero as their lowest mass state and finally
the towers with a tachyon state.

\vskip 0.5 cm
\noindent{\bf{Towers $m_n^2>0$}}

\noindent Two boundary relations
being applied
to the fields \re{solwarpKK}
lead to two linear homogeneous equations in terms of the parameters
$\sigma_n$ and $\tau_n$. The coefficients turn out to be
linear combinations of
Bessel functions evaluated for $s=0$ and $s=2\pi R$, i.e.
with arguments respectively equal to
\bea
F_0 &=& \frac{m_n}{k}
    \nonumber\\
F_2 &=& e^{2\pi k R}\frac{m_n}{k}\ .
\label{arguments}
\eea
In order to find non trivial solutions for $\sigma_n$ and $\tau_n$
and hence non trivial fields $\phi_n^{[s]}$,
the relevant determinant has to be equal to zero.
This leads to the mass equation whose solutions provide 
for chosen parameters fixing the boundary conditions
the mass eigenvalues
$m_n$ building up the related tower.
Once these mass eigenvalues are known, the corresponding
$\sigma_n/\tau_n$ ratios are deduced from one of the two original
equations.
For each set of allowed boundary conditions, both the mass equation
and a $\sigma_n/\tau_n$ relation are given in Table\re{tabletower}.

\vskip 0.5 cm
\noindent{\bf{Zero mass states}}

\noindent With two boundary relations
being applied to the fields
\re{solwarpKK0},
one obtains again two linear homogeneous equations
in terms of the parameters
$\sigma_0$ and $\tau_0$.
The condition that the relevant determinant is zero
is in fact the constraint
that has to be satisfied
by the boundary condition
parameters for a zero mass state to exist.
One of the two equations fixes again the ratio $\sigma_0/\tau_0$.

For each of the allowed set of boundary conditions, both the parameter
constraint
and a $\sigma_0/\tau_0$ equation are given in Table\re{tablezero}.
Since $e^{4\pi k R}$ takes such a large value and
since the reduced parameters
$\abar_i,\ldots$ are assumed to be of order one, approximate relations,
valid to a high degree
of precision, are easily obtained.
They are also listed in Table\re{tablezero}.
We should remark that
there is no zero mass state for the box boundary condition (Case A6).

The parameter constraint equation for a zero mass state defines a surface
in the parameter space (see approximate equation in Table\re{tablezero}).
In many cases, if one follows a path in the parameter space which crosses
the parameter constraint surface,
the lowest mass eigenvalue goes smoothly toward zero, takes the value zero
as the path goes through the surface and emerges
as a tachyon state with low $h^2=-m^2$.
Examples will be given in subsection\re{examples}
for the boundary Cases A1, A4 and A3.
A different behavior shows up in the Case A3
when in particular $\rbar_1$ passes zero.
The zero mass appears suddenly as the surface is crossed.
No small mass, no tachyons appear on either sides of the surface.
This results from the fact
that the solution $m^2=0$ is of higher order.

\vskip 0.5 cm
\noindent{\bf{Tachyon states $h^2=-m^2>0$}}

\noindent The equations for $h^2$ can simply be obtained from the equations
giving the tower masses by replacing the Bessel functions $J_n$ by
$I_n$ and $Y_n$ by $(-1)^n K_n$. They are summarized in
Table\re{tabletach} for each set of boundary conditions.

\subsection{Physical interpretation of the mass eigenvalues \label{physint}}

At this stage of the discussion, in order to make connection with physics,
one has to take into account the position
$s=s_0$ ($0\leq s_0\leq 2\pi R$)
where the
TeV brane, the brane in which we
live, is located.
Indeed the deduction of the physical masses in terms
of the mass eigenvalues depends
crucially on this position.

In our brane, the space time part of the metric
\re{metric} has a factor $e^{-2ks_0}$.
By a change of variable
\begin{equation}
\widetilde{x}_{\mu}=e^{-ks_0}x_{\mu}
\label{newvar}
\end{equation}
the space time metric in normal local units becomes $dS^2=d{\widetilde{x}}_{\mu}
d{\widetilde{x}}^{\mu}$.
Hence the physical mass $m{n,s_0}$ obeys the equation
\begin{equation}
 \widetilde{\square}_4\  \widetilde{\phi}^{[x]}_n(\widetilde{x}^{\mu})
                  =-m_{n,s_0}^2\ \widetilde{\phi}^{[x]}_n(\widetilde{x}^{\mu})\ .
\label{exactmass}
\end{equation}
Comparing \re{exactmass} and \re{boxKK} we find
\begin{equation}
m_{n,s_0}=e^{ks_0}m_n\ .
\label{massrel}
\end{equation}
This gives the relation between
the $m_n$ eigenvalues which appear in Eqs.\re{boxKK} and \re{warpKK}
and the observable physical masses $m_{n,s_0}$ in the brane.
In the case $s_0=0$, the eigenvalues
and the physical masses are equal ($m_{n,0}=m_n$).
The low lying masses
are of order 1\,TeV when $kR=6.3$. If $s_0$ differs appreciably from zero
one sees that $m_{n,s_0}$ may become large enough
to spoil the hierarchy solution. However, as will be seen later,
the solution of the problem is fully restored by an adequate increase of $kR$

\vskip 1 cm

\subsection{Examples {\label{examples}}}

In this subsection, we show for illustration the results
of some numerical computations
for Kaluza-Klein mass eigenvalue towers and field probability densities
corresponding to two sets of boundary conditions
belonging to Case A4 and Case A1 respectively.
We also make at the end some introductory comments
about the mass tower structure in the Case A3

In all the examples, the parameter $kR$ has been set equal to value $6.3$
and the parameter $\kbar$ has been chosen equal to one for convenience, so
\begin{equation}
kR=6.3\quad,\quad \kbar=1\ .
\label{param}
\end{equation}
The extension to other values of $kR$ and of $\kbar$ is outlined.

As stated in \re{physint},
the computed mass eigenvalues would be the physical masses
for a four-dimensional observer at $s_0=0$. The physical masses for $s_0\neq 0$
can be deduced from \re{massrel}.

\vskip 0.5 cm
\noindent{\bf{Case A4 mass eigenvalues}} 

\noindent The case A4 is simpler
since there is only one free parameter
fixing the boundary condition.
We remark that
the mass equation for the tower eigenvalues (see Table\re{tabletower})
is invariant under the following rescaling
with the arbitrary parameter $\lambda$
\bea
\Rbar&\rightarrow&\frac{\Rbar}{\lambda}
    \nonumber\\
\kbar&\rightarrow &\lambda \kbar
    \nonumber\\
\kabar&\rightarrow &\lambda \kabar
    \nonumber\\
m&\rightarrow& \lambda m
\label{rescalingA4}
\eea
which leaves $kR=\kbar\Rbar$ invariant.
Hence
\begin{equation}
\lambda\, m_i(\kbar,\kabar)=m_i(\lambda\kbar,\lambda\kabar)
\label{rescalingA4m1}
\end{equation}
which allows one to determine the tower mass states
for other values of $\kbar$ from the mass states corresponding
to our choice $\kbar=1$.
Indeed
\begin{equation}
m_i(\kbar,\kabar)=\kbar\, m_i\left(1,\frac{\kabar}{\kbar}\right)\ .
\label{rescalingA4m2}
\end{equation}

The ten first mass eigenvalues $m_i$  ($i=1,\dots,10$)
in the towers are given in the
Table\re{tableA4k6.3}
(remember \re{param}) and
for a few values of the boundary parameter
$\kabar$ distributed around $\kabar=4$
for which there is a zero mass state (see Table\re{tablezero}).

A few comments are worth making
\begin{itemize}

\item

Referring to the Table, one sees that the average distance
$\Delta m_i=m_{i+1}-m_i$ with $i\geq 2$ between
two consecutive states decreases very slowly along a tower and is of the order,
\begin{equation}
\Delta m_i \approx 0.25 {\rm{\ TeV}}\ .
   \label{estimation3}
\end{equation}

\item

As a result of our numerical computations ($\kbar=1$) performed
for neighbouring values of $\kbar\Rbar$, namely
\begin{equation}
\kbar\Rbar=6.3+\Delta(\kbar\Rbar)\ ,
\label{kRvalues}
\end{equation}
we have found that the corresponding mass eigenvalues
in a given tower can be deduced precisely by
\begin{equation}
m_i^{[\kbar\Rbar]}=m_i^{[6.3]}e^{-2\pi\Delta(\kbar\Rbar)}\quad{\rm{at\ fixed\ }} \kbar\ .
\label{masskR1}
\end{equation}

\item

The physical masses of the tower in the
$s=s_0$ brane, i.e. $m_{i,s_0}^{[\kbar\Rbar]}$, are obtained from \re{massrel}
(taking into account \re{param}) by
\begin{equation}
m_{i,s_0}^{[6.3]}=m_{i,0}^{[6.3]}e^{\kbar\sbar_0}
\label{masskR2}
\end{equation}
and hence, more generally
\begin{equation}
m_{i,s_0}^{[\kbar\Rbar]}
=m_{i,0}^{[6.3]}e^{\kbar\sbar_0-2\pi\Delta(\kbar\Rbar)}\ .
\label{masskR3}
\end{equation}
This is true at $\kbar$ fixed. Remember that the consequences of a change of
$\kbar$ (at fixed $\kbar\Rbar$) can be obtained from the above rescaling properties \re{rescalingA4m2}.

\item

The differences between the masses in the towers are exponentially sensitive
to the parameter $\kbar(\sbar_0-2\pi\Delta(\Rbar))$ while it is multiplicatively sensitive
to $\kbar$.
Should a Kaluza-Klein tower be discovered an approximate value of
$\kbar(\sbar_0-2\pi\Delta(\Rbar))$
could be deduced.

\item

A particular attention has to be
drawn on the first state.
Following the path in the $\kabar$ parameter space, going up
from a large negative value of $\kabar$, one sees
that the first mass eigenvalue
$m_1$ decreases faster than $m_2$ and gets equal to zero as one reaches
the surface $\kabar =4$ ($\kbar=1$)
of the zero mass constraint
(see Table\re{tablezero}).
At that point, $m_2-m_1$ is about twice the average $\Delta m_i$ value.
Once the zero mass surface is passed, the first mass eigenvalue disappears and
a tachyon state develops with $h$ increasing rapidly.

\item

The mass $m_i$ increases rather slowly when $\kabar$ decreases  toward
$-\infty$ and becomes equal to the mass $m_{i+1}$
corresponding to $\kabar=+\infty$,
exhibiting continuity of the masses as functions of $\kabar$.

\end{itemize}

\vskip 0.5 cm

\noindent{\bf{Case A1 mass eigenvalues}} 

\noindent For the Case A1,
the mass equation (see Table \re{tabletower}) is invariant under the rescaling analogous to
\re{rescalingA4}, namely
[$\Rbar\rightarrow{\Rbar}/{\lambda}$,
$\kbar\rightarrow \lambda \kbar$,
$\abar_1\rightarrow \abar_1$,
$\abar_2\rightarrow {\abar_2}/{\lambda}$,
$\abar_4\rightarrow \abar_4$,
$m\rightarrow \lambda m$],
and hence
\bea
m_i(\kbar,\abar_1,\abar_2,\abar_4)
        &=& \frac{1}{\lambda}\, m_i\left(\lambda\kbar,\abar_1,
        \frac{\abar_2}{\lambda},\abar_4\right)
    \label{rescalingA1m1}\\
m_i(\kbar,\abar_1,\abar_2,\abar_4)
        &=&\kbar\, m_i\left(1,\abar_1,
        \kbar\abar_2,\abar_4\right)\ .
\label{rescalingA1m2}
\eea

The ten lowest mass eigenvalues in the Kaluza-Klein towers are given in
Table\re{tableA1k6.3}
for an arbitrary choice
of the boundary parameters
$\abar_1=0.7$, $\abar_4=6.6286$ and for a set of values of
$\abar_2$ including the value
$\abar_2=(\abar_1\abar_4-1)/(4\abar_1\kbar)=1.3$
where the path in parameter space crosses the
zero mass surface.
The value $\abar_2=0$ is not only excluded
but appears as a singular point.
It is then convenient to
vary the values of $\abar_2$ from zero to $+\infty$
and then from $-\infty$ back to zero.
It should be noted that the passage
through the value $\abar_2=\pm \infty$ is smooth.

Apart from a few small differences,
the main structure of the mass towers is essentially the same
as for the Case A4 above.
The mass eigenvalues and the physical masses corresponding
to other values of $\Rbar$ and $\sbar_0$ (for $\kbar$ fixed) can
again be deduced by the same formulae \re{masskR1}-\re{masskR3}.
A change of $\kbar$ follows from the rescaling equation \re{rescalingA4m2}.

Moreover, comparing
the towers which have a zero mass state at their bottom
(line $\kabar=4$ for Case A4 in Table\re{tableA4k6.3}
and line $\abar_2=1.3$ for Case A1 in Table\re{tableA1k6.3}), we note that
all the masses in these two towers are practically identical.
This holds for any $kR$
in the physically allowed range.

\vskip 0.5 cm

\noindent{\bf{General comments about Case A3 mass eigenvalues}}

\noindent Summarizing

\begin{itemize}

\item The rescaling
[$\Rbar\rightarrow{\Rbar}/{\lambda}$,
$\kbar\rightarrow \lambda \kbar$,
$\rbar_1\rightarrow \lambda\rbar_1$,
$\rbar_2\rightarrow {\lambda}\rbar_2$,
$m\rightarrow \lambda m$], leads to the formulae
\bea
m_i(\kbar,\rbar_1,\rbar_2)
        &=& \frac{1}{\lambda}\, m_i\left(\lambda\kbar,\lambda\rbar_1,
        \lambda\rbar_2\right)
    \label{rescalingA3m1}\\
m_i(\kbar,\rbar_1,\rbar_2)
        &=&\kbar\, m_i\left(1,\frac{\rbar_1}{\kbar},
        \frac{\rbar_2}{\kbar}\right)\ .
\label{rescalingA3m2}
\eea

\item The approximate parameter condition
for the existence of a zero mass is (see  Table\re{tablezero})
\begin{equation}
\rbar_1(\rbar_2-4\kbar)=0 \ .
\label{A3mass0}
\end{equation}
For each of the two solutions $\rbar_1=0$ or $\rbar_2=4\kbar$,
there indeed exists a zero mass state.

\item 
For $\rbar_2=4$ (Table\re{tableA3k6.3})
the mass tower is identical to the  mass towers
with a zero mass state in the cases A1 and A4
(Tables\re{tableA1k6.3} and \re{tableA4k6.3}),
and this independently of the value of $\rbar_1$.

\item For $\rbar_1=0$,
besides the zero mass the other masses
depend on the value of $\rbar_2$.
When $\rbar_2$ moves toward $4$, the lowest non zero mass
in the tower converges also to zero.

\item When $\rbar_2$ is fixed to a given value,
all the towers corresponding to any value of $\rbar_1$ are identical,
including the tachyon if it exists (i.e. for $\rbar_2>4$).
There is of course an extra mass zero for $\rbar_1=0$.
However, for $\rbar_1$ close to zero,
neither a small mass particle nor a small $h$ tachyon appears.

\end{itemize}

\section{Physical Considerations. Probability densities}

In the context of a given boundary case,
once all the parameters are fixed and the
mass eigenvalue tower is determined, there exists
a unique field $\phi^{[s]}_n(s)$
for each mass eigenvalue
leading to a normalized
probability density field distribution $D_n(s)$
along the fifth dimension \re{scalprod}
\begin{equation}
D_n(s)=\frac{\sqrt{g}(\phi^{[s]}_n(s))^2}
        {\int_0^{2\pi R} ds\sqrt{g}(\phi^{[s]}_n(s))^2}\ .
\label{probdensity}
\end{equation}

It is convenient to
parametrize the $s$ range $[0,2\pi R]$ by the reduced variable
$x$ defined by
\begin{equation}
x=\frac{s}{2\pi R}
\label{xdefinition}
\end{equation}
with range $[0,1]$.

\vskip 0.5 cm

\noindent
{\bf{Case A1 field probability densities}}

\noindent We consider the tower labeled by
$\abar_2=1.3$ in Table\re{tableA1k6.3}.
The logarithm of the normalized
probability density for the three mass eigenvalues
$m_1=0$, $m_2=0.501\,$TeV  and $m_5=1.275\,$TeV
are given in Figure\re{figA163m0}, \re{figA163m501} and
\re{figA163m1275}
respectively as functions of $x$.
The general trends are as follows:

\begin{itemize}

\item

The probability density is a fast varying function of $x$.
In a large part of the domain the logarithm increases
or decreases linearly.

\item

A general pattern emerges. 
For $m_i$ with $i$ even,
the probability density presents around $x=0.5$
a very steep dip
down to zero as a result of a brutal but continuous change
of sign of the field at that point.

\item

Moreover, the mass $m_i$
presents $i{-}1$ probability dips in the high $x$ region
$0.95\leq x\leq 1$. 

\end{itemize}
\vskip 0.4 cm

\noindent{\bf{Relative mass eigenstate probabilities for given $x_0$}}

\noindent In a brane supposed to be
located at a certain fixed $x_0$ ($s_0=2\pi R x_0$), it is
directly possible to compare the
probabilities of the different mass eigenstates in a given tower.
Neglecting dynamical and
kinematical effects
related to the production in the available phase space, these
probabilities given in Table\re{tableA1prob1}
would account for the rate of appearance
of the mass eigenvalue states to an observer
sitting at this $x_0$. Note however that the physical masses
as seen by this observer (at $x_0\neq 0$) are not the eigenvalue masses but
vary with the $s_0$ in agreement with \re{masskR3}.

The ratios of probabilities densities are given in Table\re{tableA1prob1}
for the first ten mass eigenstates
and selected values of $x$ in its range $[0,1]$, arbitrarily normalizing to one the
highest
probability among the ten first masses considered. 

\begin{itemize}

\item

For $x$ values outside the region where
dips in the probability density appear i.e. in
the two regions
$0\leq x<0.40$ and $0.55<x<0.90$, the relative probabilities
are very weakly dependent on $x$.
In the first region $0\leq x<0.40$, the probability for $m_1$ dominates
whereas in the second region $0.55<x<0.95$, it dominates for $m_{10}$.

\item

In the dip regions  $0.45\leq x\leq 0.5$ and $0.90\leq x\leq 1$,
a chaotic behavior shows up. Small variations of $x$
may imply large fluctuations of the probabilities.

\end{itemize}

\noindent{\bf{Relative probabilities in a
given physical mass tower as a function of $x_0$}}

\noindent In Table\re{tableA1prob2},
we have taken into account the increase of $\Rbar$ dictated by \re{masskR3}
($\Rbar=6.3/(1-x_0)$)
as requested to keep the low lying physical masses exactly unchanged 
as one increases $x_0$
starting from $x_0=0$ where mass eigenvalues and physical masses coincide. 
We have limited ourselves to $\Rbar\leq 50$. 
Indeed, larger values of $\Rbar$ obviously spoil the underlying philosophy 
that all the reduced parameters have to be of order one. 
The similarity between the results presented in the two Tables 
\re{tableA1prob1} and \re{tableA1prob2}
should be noted though the physical interpretation is widely different.

\vskip 0.5 cm

\noindent
{\bf{Case A4 field probability densities}}

\noindent Summarizing

\begin{itemize}

\item

Figures\re{figA463m0} and \re{figA463m766}
show the logarithm
of the normalized probability density as a function of $x$ for $\kabar=4$,
respectively for two mass eigenvalues chosen for illustration $m_1=0$ and
$m_3=0.766\,$TeV.

\item

All the probabilities are seen to increase very fast with $x$
since the logarithm is essentially
a linear function for $0.1\leq x\leq 0.9$.

\item

Due to the boundary condition $\phi(0)=0$ (Table\re{tableBCdeux})
there is a sharp dip for $s=0$.

\item

In the high $x$ region ($0.9\leq x \leq 1$)
the probability for the mass
$m_i$ exhibits $i{-}1$ sharp dips corresponding to zeros in the field.

\item

In the whole region $0.1\leq x\leq 0.9$
the relative probabilities
in a tower are very close to those of Case A1 for $x$
between $0.5$ and $0.9$ (Table \re{tableA1prob1}).

\item 

In the two extreme regions (where there are dips)
the relative probabilities
exhibit a chaotic behavior as
in the dip regions for A1.

\item 

The same considerations hold as for the case A1 regarding the requested
readjustment of $\Rbar$ to keep the physical masses unchanged when the position
$x_0$ is changed.

\end{itemize}

\section{Conclusions}

Inspired by the warped five-dimensional scenario of
Randall and Sundrum and
restricting to the case of a real massless scalar field
supposed to propagate in the bulk,
we have developed a similar warped model,
keeping all the basic parameters adjusted
in terms of the Planck mass as the only dimensionful scale.
This, in the end, solves the mass scale hierarchy problem.

We have concentrated primarily on a careful study of the hermiticity
(symmetry, self-adjointness) and commutativity
of the operators susceptible to be used
to validly establish the Kaluza-Klein reduction equations
in the five dimensional warped space. Postulating
that the fifth extra dimension $s$
is compactified on a strip $0\leq s\leq 2\pi R$ and that the metric
has no dicontinuity, we have enumerated
all the allowed boundary conditions.
From them we have deduced all the Kaluza-Klein towers mass equations 
providing the mass eigenvalues,
as well as the tachyon mass equations.

We have discussed how these mass eigenstates show up with
physical masses depending on the location of our brane on the $s$ axis. 

As an illustration,
we have carried on some numerical computations
for the three sets of boundary conditions
A1, A4 and A3 in order to visualize the structure of the towers
and to investigate their main properties.
The other cases can be studied along the same lines.

The structure of the eigenvalue towers depends generally
in a sensitive way on the value of the basic parameter $kR$
and to a smaller extend on the boundary parameters
and on the reduced parameter $\kbar$.
Apart from small differences,
the main structure is the same in the three Cases considered.

One notices that
the first tower masses are of the order TeV for $kR$
around 6.3 as expected for solving the mass
hierarchy problem.

In general, for specific values of the boundary parameters
there exists a zero mass state.
For parameters close to these values the first mass state in the tower
is often either a particle with a small mass or a tachyon.

One observes that the mass spacing (discarding the first mass state)
is very stable
within a given tower and is exponentially sensitive to the value of $k(s_0-2\pi R)$ where
$s_0$ is the position of our brane.
Hence should a Kaluza-Klein tower be
observed experimentally, a good estimation of this basic parameter would result.

The normalized field probability density for any physical mass state 
in a tower
can easily be computed for fixed values of the boundary parameters
as a function of  $\kbar$, $\Rbar$ and $x_0$.
Neglecting dynamical and kinematical effects the ratios of the probabilities
among the first masses in a given tower
evaluated at $x_0$ would express the relative
intensities of the eventually observed mass peaks.

\vskip 0.5 cm

\vskip 1 cm

\noindent{\bf{Acknowledgment}}

The authors are grateful to Professor David Fairlie
for a discussion.

\newpage

\appendix

\section{Appendix: The Bessel Functions. Notations}{\label{appendixA}}

The Bessel function $J_p$, as well as $Y_p$, satisfies
\bea
\biggl(y^2\partial_y^2 + y \partial_y +\left(y^2-p^2\right)\biggr)J_p(y)&=&0
       \nonumber\\
\partial_y J_p(y)+p \frac{J_p(y)}{y}-J_{p-1}(y)&=&0
       \nonumber\\
J_p(y)-2(p-1)\frac{J_{p-1}(y)}{y}+ J_{p-2}(y)&=&0\ .
     \label{bessel}
\eea
A useful identity is
\begin{equation}
J_2(y)Y_1(y)-J_1(y)Y_2(y)-\frac{2}{\pi y}=0\ .
     \label{bessel1}
\end{equation}
The modified Bessel function $I_p$, as well as $K_p$, satisfies
\be
\biggl(y^2\partial_y^2 + y \partial_y -\left(y^2+p^2\right)\biggr)I_p(y)=0
     \label{besselmod1}
\end{equation}
and (note the sign differences)
\bea
\partial_y I_p(y)+p \frac{I_p(y)}{y}-I_{p-1}(y)&=&0
      \nonumber\\
\partial_y K_p(y)+p \frac{K_p(y)}{y}+K_{p-1}(y)&=&0
      \nonumber\\
I_p(y)+2(p-1)\frac{I_{p-1}(y)}{y}-I_{p-2}(y) &=&0
      \nonumber\\
K_p(y)-2(p-1)\frac{K_{p-1}(y)}{y}-K_{p-2}(y) &=&0\ .
      \label{besselmod2}
\eea
The corresponding useful identity is
\begin{equation}
I_2(y)K_1(y)+I_1(y)K_2(y)-\frac{1}{y}=0\ .
     \label{besselmod3}
\end{equation}

\section{Appendix: Singularities{\label{appendixB}}}

In the main part of the paper, we have analyzed the case of a warped space \re{metric}
induced by a cosmological constant \re{klambda} with a fifth dimension compactified to a
strip
$0\leq s\leq 2\pi R$.
We made the choice of a positive $k$ 
everywhere in the space
and postulated that there was no singularity.

First, the $k$ negative case is simply related
to the positive case by the exchange
$s\leftrightarrow {-}s{+}2\pi R$ which hence makes the sign of $k$ an arbitrary choice.

However, it may be assumed that in some region of $s$ the constant $k$
is positive and in another region it is negative.
With either $k$ positive or negative there must be at least a singular point
$s_s$ where a transition in the metric occurs. Writing
\bea
{\rm{for\ }} s<s_s&\quad& dS^2=Ce^{2k(s-s_s)} dx_{\mu}dx^{\mu}-ds^2
     \nonumber\\
{\rm{for\ }} s=s_s&\quad& dS^2=\ \ C\quad dx_{\mu}dx^{\mu}-ds^2
     \nonumber\\
{\rm{for\ }} s>s_s&\quad& dS^2=Ce^{-2k(s-s_s)} dx_{\mu}dx^{\mu}-ds^2 
\label{singular}
\eea
the metric is continuous at $s=s_s$ as it should but
its first derivative has a discontinuity $\pm 4kC$ and
its second derivative a $\delta$-function behavior.
In principle, there could be any finite number of such singularities.

If there is an even number of singularities ($2m,m\geq 1$), the strip compactification
can be transformed to an orbifold compactification by identifying the edges
$0$ and $2\pi R$ closing the strip to a circle, 
allowing periodic or antiperiodic conditions. Without loss of generality,
one may chose these $2m$ singularities to be located at
$0,s_1,s_2,\ldots,s_{2m-1}$ with $0<s_1<s_2<\ldots<s_{2m-1}<2\pi R$.
A necessary condition for the closure is
\begin{equation}
\sum_{i=1}^{m}s_{2i-1}-\sum_{i=1}^{m-1}s_{2i}=\pi R \ .
\label{condorbi}
\end{equation}
Indeed the total length of the region where $k$ is positive
must be equal to the total length where $k$ is negative and
must
thus be equal to one half of the length $2\pi R$ of the circle.

We expect to come back to the problem of the warped Kaluza-Klein
towers with singularities in a forthcoming paper.
This change of metric has a direct impact
on the establishment of the boundary conditions
and the treatment of the Kaluza-Klein equations.

 \newpage
\vskip 0.5 cm
\begin{table}
\caption{Two boundary conditions
{\label{tableBCdeux}}
}
\vspace{1 cm}
\hspace{1 cm}
\scriptsize
{
\begin{tabular}{|l|l|l|}
\hline
\multicolumn{3}{|c|}
      {Two Boundary Conditions}
        \\ \hline
    Case&Boundary Conditions&Reduced Parameters
      \\ \hline\hline
    A1              & $\phi(2\pi R)=e^{4\pi k R}\left(\alpha_1\phi(0)
                     +\alpha_2\partial_s\phi(0)\right) $
                    &$\alpha_1=\abar_1\quad,
                    \quad \alpha_2=\abar_2/M_{Pl}\neq 0$
                       \\
                    & $\partial_s\phi(2\pi R)=e^{4\pi k R}
                    \left(\frac{\alpha_1\alpha_4-1}
                    {\alpha_2}\phi(0)
                    +\alpha_4\partial_s\phi(0)\right) $
                    &$\alpha_4=\abar_4$
       \\ \hline
    A2              & $\phi(2\pi R)=e^{4\pi k R}\alpha_1\phi(0)$
                    &$\alpha_1=\abar_1\neq 0$
                       \\
                    & $\partial_s\phi(2\pi R)=e^{4\pi k R}
                    \left(\alpha_3\phi(0)
                    +\frac{1}{\alpha_1}\partial_s\phi(0)\right) $
                    &$\alpha_3=\abar_3\,M_{Pl}$
       \\ \hline
    A3              & $\partial_s\phi(0)=\rho_1\phi(0) $
                    &$\rho_1=\rbar_1\,M_{Pl}$
                       \\
                    & $\partial_s\phi(2\pi R)=\rho_2\phi(2\pi R) $
                    &$\rho_2=\rbar_2\,M_{Pl}$
       \\ \hline
    A4              & $\phi(0)=0 $
                     &
                       \\
                    & $\partial_s\phi(2\pi R)=\kappa\phi(2\pi R) $
                    &$\kappa=\kabar\,M_{Pl}$
       \\ \hline
    A5              & $\phi(2\pi R)=0 $
                     &
                       \\
                    & $\partial_s\phi(0)=\zeta\phi(0) $
                    &$\zeta=\zbar\,M_{Pl}$
       \\ \hline
    A6              & $\phi(0)=0 $
                     &
                       \\
                    & $\phi(2\pi R)=0 $
                     &
      \\ \hline
\end{tabular}
   }
\end{table}

\vfill\break
\newpage

\newpage
\vskip 0.5 cm
\begin{table}
\caption{Three boundary conditions
{\label{tableBCtrois}}
}
\vspace{1 cm}
\hspace{1 cm}
\scriptsize
{
\begin{tabular}{|l|l|l|}
\hline
\multicolumn{3}{|c|}
      {Three Boundary Conditions}
        \\ \hline
    Case&Boundary Conditions&Reduced Parameters
      \\ \hline\hline
    B1              & $\phi(2\pi R)=e^{4\pi k R}\lambda_1\phi(0)$
                    &$\lambda_1=\lbar_1$
                       \\
                    & $\partial_s\phi(0)=\lambda_2\phi(0)$
                    &$\lambda_2=\lbar_2 M_{Pl}$
                       \\
                    & $\partial_s\phi(2\pi R)=
                    e^{4\pi k R}\lambda_3\phi(0)$
                    &$\lambda_3=\lbar_3M_{Pl}$
       \\ \hline
    B2              & $\phi(0)=0$
                    &$$
                       \\
                    & $\phi(2\pi R)=e^{4\pi k R}\mu_1\partial_s\phi(0)$
                    &$\mu_1=\mbar_1/ M_{Pl}$
                       \\
                    & $\partial_s\phi(2\pi R)=e^{4\pi k R}\mu_2
                    \partial_s\phi(0)$
                    &$\mu_2=\mbar_2$
       \\ \hline
    B3              & $\phi(0)=0$
                    &$$
                       \\
                    & $\partial_s\phi(0)=0$
                    &$$
                       \\
                    & $\partial_s\phi(2\pi R)=\nu\phi(2\pi R)$
                    &$\nu=\nbar M_{Pl}$
       \\ \hline
    B4              & $\phi(0)=0$
                    &$$
                       \\
                    & $\phi(2\pi R)=0$
                    &$$
                       \\
                    & $\partial_s\phi(0)=0$
                    &$$

      \\ \hline
\end{tabular}
   }
\end{table}
\vfill\break
\newpage

\begin{table}
{\caption{Two boundary conditions : zero mass state constraints
($\rightarrow\ :$ approximate
relation resulting from $e^{4k\pi R}$
being very large){\label{tablezero}}}}
\vspace{1 cm}
\hspace{1 cm}
\scriptsize
{
\begin{tabular}{|l|l|l|}
\hline
\multicolumn{3}{|c|}
      {Parameter constraints for zero mass states}
        \\ \hline
    Case&Parameter constraint&$\sigma_0$, $\tau_0$ relation
      \\ \hline\hline
    A1              & $e^{8\pi k R}\left(4\alpha_1\alpha_2 k
                     -\alpha_1\alpha_4 +1\right)
                     $
                    &$e^{4\pi k R}\left(e^{4\pi k R}
                     -\alpha_1-4\alpha_2 k\right)\sigma_0
                    =\left(e^{4\pi k R}\alpha_1-1\right)\tau_0$
           \\
                    & $\quad -8\ e^{4\pi k R}\alpha_2 k
                     $
                     &
            \\
                     & $\quad\quad +\left(\alpha_1\alpha_4
                     +4\alpha_2\alpha_4 k -1\right)=0
                     $
                     &
            \\
                    \cline{2-3}
                     &$\rightarrow \quad 4\alpha_1\alpha_2 k
                     -\alpha_1\alpha_4 +1\approx 0$
                     &$\rightarrow \quad \sigma_0 \approx 0$
       \\ \hline\hline
    A2              & $e^{8\pi k R}\left(4\alpha_1^2 k
                    -\alpha_1\alpha_3 \right)
                       $
                    &$e^{4\pi k R}\left(e^{4\pi k R}
                    -\alpha_1\right)\sigma_0
                    =\left(e^{4\pi k R}\alpha_1-1\right)\tau_0$
                       \\
                       & $\quad -8\ e^{4\pi k R}\alpha_1 k
                       $
                       &
                       \\
                       & $\quad\quad+\left(\alpha_1\alpha_3+4 k \right)=0
                       $
                       &
            \\
                    \cline{2-3}
                     &$\rightarrow \quad 4\alpha_1^2k-\alpha_3\approx 0$
                     &$\rightarrow \quad \sigma_0 \approx 0$
       \\ \hline\hline
    A3              & $e^{8\pi k R}\rho_1\left(4 k-\rho_2 \right)
                       +\rho_2\left(\rho_1-4 k \right)=0 $
                    &$\left(4 k-\rho_1\right)\sigma_0=\rho_1\tau_0$
            \\
                    \cline{2-3}
                     &$\rightarrow \quad \rho_1\left(\rho_2-4 k\right)
                     \approx 0$
                     &$\rightarrow \quad \left(4 k-\rho_1\right)\sigma_0
                     =\rho_1\tau_0 $
       \\ \hline\hline
    A4              & $ e^{8\pi k R}\left(4 k-\kappa \right)+\kappa =0 $
                     &$\sigma_0=-\tau_0$
            \\
                    \cline{2-3}
                     &$\rightarrow \quad \kappa - 4 k \approx 0$
                     &$\rightarrow \quad \sigma_0=-\tau_0 $
       \\ \hline\hline
    A5              & $e^{8\pi k R}\zeta +\left(4k-\zeta\right) =0$
                     & $e^{8\pi k R}\sigma_0=-\tau_0$
            \\
                    \cline{2-3}
                     &$\rightarrow \quad \zeta\approx 0$
                     &$\rightarrow \quad\sigma_0\approx 0$
       \\ \hline\hline
    A6              & No zero mass state
                     & $\sigma_0=\tau_0=0$
       \\ \hline
\end{tabular}
   }
\end{table}

\vfill\break
\newpage

\begin{table}
\caption{Mass tower equations for two boundary conditions
{\label{tabletower}}
}
\vspace{1 cm}
\hspace{1 cm}
\tiny
{
\begin{tabular}{|c|l|l|}
\hline
\multicolumn{3}{|c|}
      {Notations: $E=e^{2\pi k R}$, $F_0=\frac{m}{k}$,
      $F_2=e^{2\pi k R}\frac{m}{k}$
      {\phantom{$\biggl.\biggr.$}}
       }
   \\ \hline
      {\phantom{$\biggl.\biggr.$}}
Case
      & Mass equation
      & $\sigma, \tau$ relation
     \\  \hline
      {\phantom{$\biggl.\biggr.$}}
  A1
       &
    $\alpha_2 m E\bigl[ \left(J_2(F_0)Y_1(F_2)
    -J_1(F_2)Y_2(F_0)\right)\alpha_1\bigr.$
       &
    $\left(J_2(F_0)\alpha_1+J_1(F_0)\alpha_2 m
    - J_2(F_2)\right) \sigma   $
      \\
      {\phantom{$\biggl.\biggr.$}}
       &
    $\quad \quad + \biggl.\alpha_2 m \left(J_1(F_0)Y_1(F_2)
    -J_1(F_2)Y_1(F_0)\right)\bigr]$
       &
    $\quad \quad + \left(Y_2(F_0)\alpha_1+Y_1(F_0)\alpha_2 m
    - Y_2(F_2)\right)\tau=0$
       \\
      {\phantom{$\biggl.\biggr.$}}
       &
    $+ \bigl[\left(\alpha_1\alpha_4-1\right)
    \left(J_2(F_2)Y_2(F_0)-J_2(F_0)Y_2(F_2)
    \right)
                         \bigr.$
       &
       \\
      {\phantom{$\biggl.\biggr.$}}
       &
    $\quad \quad + \alpha_2\alpha_4 m\left(J_2(F_2)Y_1(F_0)
    -J_1(F_0)Y_2(F_2)\right)$
       &
       \\
      {\phantom{$\biggl.\biggr.$}}
       &
    $\quad \quad - \left.\alpha_2 \frac{4k}{\pi} \right]=0$
       &
      \\ \hline
      {\phantom{$\biggl.\biggr.$}}
  A2
       &
    $ m E\alpha_1^2\bigl[J_2(F_0)Y_1(F_2)-J_1(F_2)Y_2(F_0)\bigr]$
       &
    $\left(J_2(F_0)\alpha_1- J_2(F_2)\right) \sigma   $
      \\
      {\phantom{$\biggl.\biggr.$}}
       &
    $+ \bigl[ \left(J_2(F_2)Y_2(F_0)-J_2(F_0)Y_2(F_2)\right)
                         \alpha_1\alpha_3\bigr. $
       &
    $\quad \quad + \left(Y_2(F_0)\alpha_1 - Y_2(F_2)\right)\tau=0$
       \\
     {\phantom{$\biggl.\biggr.$}}
       &
    $\quad \quad - \alpha_1 \frac{4k}{\pi}$
       &
       \\
      {\phantom{$\biggl.\biggr.$}}
       &
    $\quad \quad + \bigl.m\left(J_2(F_2)Y_1(F_0)
    -J_1(F_0)Y_2(F_2)\right)\bigr]=0$
       &
      \\ \hline
      {\phantom{$\biggl.\biggr.$}}
  A3
       &
    $m E\bigl[ \rho_1\left(J_1(F_2)Y_2(F_0)
    -J_2(F_0)Y_1(F_2)\right)\bigr.$
       &
    $\left(J_1(F_0) m - J_2(F_0)\rho_1\right) \sigma   $
      \\
      {\phantom{$\biggl.\biggr.$}}
       &
    $\quad \quad + \bigl.m\left(J_1(F_0)Y_1(F_2)
    -J_1(F_2)Y_1(F_0)\right)\bigr]$
       &
    $\quad \quad + \left(Y_1(F_0) m - Y_2(F_0)\rho_1\right)\tau=0$
       \\
      {\phantom{$\biggl.\biggr.$}}
       &
    $ + \rho_2 \bigl[ \rho_1\left(J_2(F_0)Y_2(F_2)
    -J_2(F_2)Y_2(F_0)\right)\bigr.$
       &
       \\
      {\phantom{$\biggl.\biggr.$}}
       &
    $\quad \bigl. \quad +  m\left(J_2(F_2)Y_1(F_0)
    -J_1(F_0)Y_2(F_2)\right)\bigr]=0$
       &
      \\ \hline
      {\phantom{$\biggl.\biggr.$}}
  A4
       &
    $m E\bigl( J_1(F_2)Y_2(F_0)-J_2(F_0)Y_1(F_2)\bigr)$
       &
    $J_2(F_0) \sigma+Y_2(F_0)\tau=0   $
      \\
      {\phantom{$\biggl.\biggr.$}}
       &
    $\quad \quad + \kappa \bigl(J_2(F_0)Y_2(F_2)
    -J_2(F_2)Y_2(F_0)\bigr)=0$
       &
    $\quad $
      \\ \hline
      {\phantom{$\biggl.\biggr.$}}
  A5
       &
    $\zeta\bigl( J_2(F_2)Y_2(F_0)-J_2(F_0)Y_2(F_2)\bigr)$
       &
    $J_2(F_2) \sigma +Y_2(F_2)\tau=0   $
      \\
      {\phantom{$\biggl.\biggr.$}}
       &
    $\quad +m \bigl(J_1(F_0)Y_2(F_2)-J_2(F_2)Y_1(F_0)\bigr)=0$
       &
    $\quad $
      \\ \hline
      {\phantom{$\biggl.\biggr.$}}
  A6
       &
    $ J_2(F_2)Y_2(F_0)-J_2(F_0)Y_2(F_2)=0$
       &
    $J_2(F_0) \sigma +Y_2(F_0)\tau=0   $
      \\
      {\phantom{$\biggl.\biggr.$}}
       &
    $$
       &
    $\quad $
    \\ \hline
\end{tabular}
        }
\end{table}

\begin{table}
\caption{Tachyon equations for two boundary conditions
{\label{tabletach}}
}
\vspace{1 cm}
\hspace{1 cm}
\tiny
{
\begin{tabular}{|c|l|l|}
\hline
\multicolumn{3}{|c|}
      {Notations: $E=e^{2\pi k R}$, $F_0=\frac{h}{k}$,
      $F_2=e^{2\pi k R}\frac{h}{k}$ ($m^2=-h^2$)
      {\phantom{$\biggl.\biggr.$}}
       }
   \\ \hline
      {\phantom{$\biggl.\biggr.$}}
Case
      & Mass equation
      & $\sigma, \tau$ relation
     \\  \hline
      {\phantom{$\biggl.\biggr.$}}
  A1
       &
    $\alpha_2 h E\bigl[ \left(-I_2(F_0)K_1(F_2)
    -I_1(F_2)K_2(F_0)\right)\alpha_1\bigr.$
       &
    $\left(I_2(F_0)\alpha_1+I_1(F_0)\alpha_2 h
    - I_2(F_2)\right) \sigma   $
      \\
      {\phantom{$\biggl.\biggr.$}}
       &
    $\quad \quad + \biggl.\alpha_2 h \left(-I_1(F_0)K_1(F_2)
    +I_1(F_2)K_1(F_0)\right)
    \biggr]$
       &
    $\quad \quad + \left(K_2(F_0)\alpha_1-K_1(F_0)\alpha_2 h
    - K_2(F_2)\right)\tau=0$
       \\
      {\phantom{$\biggl.\biggr.$}}
       &
    $+ \bigl[\left(\alpha_1\alpha_4-1\right) \left(I_2(F_2)K_2(F_0)
    -I_2(F_0)K_2(F_2)
    \right)
                         \bigr.$
       &
       \\
      {\phantom{$\biggl.\biggr.$}}
       &
    $\quad \quad + \alpha_2\alpha_4 h\left(-I_2(F_2)K_1(F_0)
    -I_1(F_0)K_2(F_2)\right)$
       &
       \\
      {\phantom{$\biggl.\biggr.$}}
       &
    $\quad \quad + \left.2\alpha_2 k \right]=0$
       &
      \\ \hline
      {\phantom{$\biggl.\biggr.$}}
  A2
       &
    $\alpha_1^2 h E\bigl[ \left(-I_2(F_0)K_1(F_2)
    -I_1(F_2)K_2(F_0)\right)\bigr.$
       &
    $\left(I_2(F_0)\alpha_1- I_2(F_2)\right) \sigma   $
      \\
      {\phantom{$\biggl.\biggr.$}}
       &
    $+ \bigl[ \left(I_2(F_2)K_2(F_0)-I_2(F_0)K_2(F_2)\right)
                         \alpha_1\alpha_3\bigr. $
       &
    $\quad \quad + \left(K_2(F_0)\alpha_1 - K_2(F_2)\right)\tau=0$
       \\
      {\phantom{$\biggl.\biggr.$}}
       &
    $\quad \quad + h\left(-I_2(F_2)K_1(F_0)-I_1(F_0)K_2(F_2)\right)$
       &
       \\
      {\phantom{$\biggl.\biggr.$}}
       &
    $\quad \quad + \bigl.2\alpha_1 k\bigr]=0$
       &
      \\ \hline
      {\phantom{$\biggl.\biggr.$}}
  A3
       &
    $h E\bigl[ \rho_1\left(I_1(F_2)K_2(F_0)
    +I_2(F_0)K_1(F_2)\right)\bigr.$
       &
    $\left(I_1(F_0) h - I_2(F_0)\rho_1\right) \sigma   $
      \\
      {\phantom{$\biggl.\biggr.$}}
       &
    $\quad \quad + \bigl.h\left(-I_1(F_0)K_1(F_2)
    +I_1(F_2)K_1(F_0)\right)\bigr]$
       &
    $\quad \quad - \left(K_1(F_0) h + K_2(F_0)\rho_1\right)\tau=0$
       \\
      {\phantom{$\biggl.\biggr.$}}
       &
    $ + \rho_2 \bigl[ \rho_1\left(I_2(F_0)K_2(F_2)
    -I_2(F_2)K_2(F_0)\right)\bigr.$
       &
       \\
      {\phantom{$\biggl.\biggr.$}}
       &
    $\quad \bigl. \quad -h\left(I_2(F_2)K_1(F_0)
    + I_1(F_0)K_2(F_2)\right)\bigr]=0$
       &
      \\ \hline
      {\phantom{$\biggl.\biggr.$}}
  A4
       &
    $h E\bigl( I_1(F_2)K_2(F_0)+I_2(F_0)K_1(F_2)\bigr)$
       &
    $I_2(F_0) \sigma+K_2(F_0)\tau=0   $
      \\
      {\phantom{$\biggl.\biggr.$}}
       &
    $\quad \quad + \kappa \bigl(I_2(F_0)K_2(F_2)
    -I_2(F_2)K_2(F_0)\bigr)=0$
       &
    $\quad $
      \\ \hline
      {\phantom{$\biggl.\biggr.$}}
  A5
       &
    $\zeta\bigl( I_2(F_2)K_2(F_0)-I_2(F_0)K_2(F_2)\bigr)$
       &
    $I_2(F_2) \sigma +K_2(F_2)\tau=0   $
      \\
      {\phantom{$\biggl.\biggr.$}}
       &
    $\quad +h \bigl(I_1(F_0)K_2(F_2)+I_2(F_2)K_1(F_0)\bigr)=0$
       &
    $\quad $
      \\ \hline
      {\phantom{$\biggl.\biggr.$}}
  A6
       &
    $ I_2(F_2)K_2(F_0)-I_2(F_0)K_2(F_2)=0$
       &
    $I_2(F_0) \sigma +K_2(F_0)\tau=0   $
      \\
      {\phantom{$\biggl.\biggr.$}}
       &
    $$
       &
    $\quad $
    \\ \hline
\end{tabular}
        }
\end{table}

\newpage

\begin{table}
\caption{
{\label{tableA4k6.3}}
}\vspace{1 cm}
\hspace{1 cm}
\scriptsize
{
\begin{tabular}{|c|c|c|c|c|c|c|c|c|c|c|c|}
\hline
\multicolumn{12}{|c|}
      {Towers of mass eigenvalues for the Case A4, masses are in TeV,
      $\kbar\Rbar=6.3$, $\kbar=1$
      $\phantom{\Biggl[\Biggr]}$}
        \\ \hline
    $\kabar$&h&$m_1$&$m_2$&$m_3$&$m_4$&$m_5$&$m_6$&$m_7$&$m_8$&$m_9$&$m_{10}$
      \\ \hline\hline
 - 100.0& &0.4&0.655&0.9048&1.152&1.398&1.644&1.889&2.134&2.379&2.624
      \\ \hline
 - 12.0& &0.3769&0.62&0.8599&1.1&1.34&1.582&1.825&2.068&2.312&2.557
      \\ \hline
 - 8.0& &0.3672&0.6072&0.846&1.086&1.328&1.571&1.814&2.058&2.303&2.548
      \\ \hline
 - 4.0& &0.349&0.5866&0.8265&1.069&1.312&1.557&1.802&2.047&2.293&2.539
       \\ \hline
    0.0& &0.301&0.5512&0.7993&1.047&1.294&1.542&1.789&2.036&2.283&2.529
        \\ \hline
  3.9& &0.0605&0.5025&0.768&1.024&1.276&1.526&1.775&2.024&2.272&2.52
        \\ \hline
 3.99& &0.0192&0.5014&0.766&1.023&1.275&1.525&1.774&2.023&2.272&2.52
        \\ \hline
 3.999& &0.00609&0.5016&0.766&1.023&1.275&1.525&1.775&2.023&2.272&2.52
        \\ \hline
4.0&  & 0&0.5013&0.766&1.023&1.275&1.525&1.775&2.023&2.272&2.52
        \\ \hline
4.001& 0.00609 & &0.5016&0.766&1.023&1.275&1.525&1.775&2.023&2.272&2.52
        \\ \hline
4.01& 0.01926& &0.5016&0.766&1.023&1.275&1.525&1.775&2.023&2.272&2.52
        \\ \hline
4.1& 0.0612& &0.501&0.766&1.022&1.275&1.525&1.774&2.023&2.271&2.519
        \\ \hline
8.0& 0.4826& &0.4624&0.7380&1.0&1.256&1.51&1.761&2.011&2.261&2.51
        \\ \hline
12.0& 0.8091& &0.4428&0.718&0.9819&1.241&1.496&1.749&2.0&2.251&2.501
        \\ \hline
16.0& 1.128& &0.4327&0.7058&0.9689&1.228&1.484&1.737&1.99&2.241&2.492
        \\ \hline
20.0& 1.445& &0.426&0.6969&0.959&1.218&1.474&1.728&1.981&2.233&2.484
        \\ \hline
100.0& 7.737& &0.408&0.6689&0.9227&1.175&1.426&1.676&1.926&2.176&2.426
        \\ \hline
\end{tabular}
   }
\end{table}
\vfill\break

\begin{table}
\caption{
{\label{tableA1k6.3}}
}
\vspace{1 cm}
\hspace{1 cm}
\scriptsize
{
\begin{tabular}{|c|c|c|c|c|c|c|c|c|c|c|c|}
\hline
\multicolumn{12}{|c|}
      {Towers of mass eigenvalues for the Case A1, masses are in TeV, $\kbar\Rbar=6.3$,
      $\kbar=1$, $\abar_1=0.7$, $\abar_4=6.6286$
      $\phantom{\Biggl[\Biggr]}$}
        \\ \hline
  $\abar_2$&h&$m_1$&$m_2$&$m_3$&$m_4$&$m_5$&$m_6$&$m_7$&$m_8$&$m_9$&$m_{10}$
      \\ \hline\hline
0.& $\infty$ & &0.4035&0.6613&0.9129&1.162&1.411&1.659&1.907&2.154&2.402
\\ \hline
0.02& 20.31 & &0.4051&0.6639&0.9165&1.167&1.416&1.665&1.914&2.163&2.411
\\ \hline
0.2& 1.918 & &0.4204&0.6884&0.949&1.206&1.462&1.716&1.969&2.222&2.473
\\ \hline
1.& 0.2269 & &0.4873&0.7575&1.015&1.269&1.521&1.771&2.019&2.268&2.516
\\ \hline
1.29&0.0340 & &0.5012&0.766&1.022&1.274&1.525&1.774&2.023&2.271&2.519
\\ \hline
1.299& 0.0107 & &0.5013&0.7669&1.022&1.275&1.525&1.774&2.023&2.271&2.519
\\ \hline
1.3&   & 0&0.5013&0.766&1.023&1.275&1.525&1.774&2.023&2.271&2.519
\\ \hline
1.301&  & 0.01064&0.5013&0.766&1.023&1.275&1.525&1.774&2.023&2.271&2.519
\\ \hline
1.31& &0.03359&0.5017&0.766&1.023&1.275&1.525&1.774&2.023&2.271&2.519
\\ \hline
1.6& &0.159&0.5112&0.773&1.027&1.278&1.528&1.777&2.025&2.273&2.521
\\ \hline
2.6& &0.24&0.5269&0.7833&1.035&1.284&1.533&1.781&2.029&2.277&2.524
\\ \hline
40.0& &0.2984&0.5497&0.7983&1.046&1.293&1.541&1.788&2.035&2.282&2.529
\\ \hline
100.0& &0.3&0.5511&0.7989&1.046&1.294&1.541&1.788&2.035&2.282&2.529
\\ \hline
$\pm\infty$&  &0.3010&0.5512&0.7993&1.047&1.294&1.541&1.788&2.035&2.282&2.529
\\ \hline
 - 100.0&   &0.3021&0.5518&0.7997&1.047&1.294&1.541&1.788&2.035&2.282& 2.529
\\ \hline
 - 40.0&   &0.3037&0.5527&0.8011&1.048&1.295&1.542&1.789&2.035&2.282& 2.529
\\ \hline
 - 2.0&   &0.3375   &0.5763&0.8178&1.061&1306.&1.551&1.797&2.043&2.289& 2.535
\\ \hline
 - 0.001&   &0.4034 &0.6612&0.9127&1.162&1.411&1.659&1.906&2.154&2.402& 2.648
\\ \hline
\end{tabular}
   }

\end{table}

\begin{table}
\caption{
{\label{tableA3k6.3}}
}
\vspace{1 cm}
\hspace{1 cm}
\scriptsize
{
\begin{tabular}{|c|c|c|c|c|c|c|c|c|c|c|c|}
\hline
\multicolumn{12}{|c|}
      {Towers of mass eigenvalues for the Case A3, masses are in TeV,
      $\kbar\Rbar=6.3$, $\kbar=1$
      $\phantom{\Biggl[\Biggr]}$}
        \\ \hline
  $\rbar_2$&h&$m_1$&$m_2$&$m_3$&$m_4$&$m_5$&$m_6$&$m_7$&$m_8$&$m_9$&$m_{10}$
      \\ \hline\hline
 - 4000.0& &0.4034&0.6611&0.9127&1.162&1.411&1.659&1.906&2.154&2.401&2.648
 \\ \hline
 - 400.0 & &0.4025&0.6596&0.9106&1.160&1.407&1.655&1.902&2.149&2.396&2.642
 \\ \hline
 - 40.0  & &0.3940&0.6459&0.8919&1.136&1.380&1.623&1.866&2.109&2.352&2.596
 \\ \hline
 - 4.0   & &0.3487&0.5866&0.8265&1.069&1.312&1.557&1.802&2.047&2.293&2.539
 \\ \hline
0.0      & &0.3010&0.5512&0.7993&1.047&1.294&1.541&1.788&2.035&2.282&2.529
\\ \hline
3.0      & &0.1807&0.5139&0.7750&1.029&1.279&1.529&1.778&2.026&2.274&2.522
\\ \hline
3.9      & &0.0605&0.5025&0.766&1.023&1.275&1.525&1.775&2.023&2.271&2.519
\\ \hline
3.99     & &0.0192&0.5014&0.766&1.023&1.275&1.525&1.774&2.023&2.271&2.519
\\ \hline
4.0      & &0.    &0.5012&0.766&1.023&1.275&1.525&1.774&2.023&2.271&2.519
\\ \hline
4.01&0.0193&      &0.5011&0.766&1.023&1.275&1.525&1.774&2.023&2.271&2.519
\\ \hline
4.1 &0.0612&      &0.5000&0.766&1.022&1.274&1.524&1.774&2.023&2.271&2.519
\\ \hline
5.0 &0.2046&      &0.4895&0.7590&1.017&1.270&1.521&1.771&2.020&2.268&2.517
\\ \hline
8.0 &0.4826&      &0.4624&0.7380&1.0  &1.256&1.509&1.761&2.011&2.260&2.509
\\ \hline
12.0&0.8092&      &0.4425&0.7180&0.9818&1.240&1.495&1.748&2.0 &2.250&2.500
\\ \hline
40.0&3.021 &      &0.4142&0.6786&0.9365&1.192&1.446&1.699&1.952&2.204&2.455
\\ \hline
400.0&31.31&      &0.4045&0.6629&0.9152&1.165&1.415&1.663&1.912&2.160&2.408
\\ \hline
4000.0&314.1&     &0.4036&0.6614&0.9131&1.163&1.411&1.659&1.907&2.155&2.402
\\ \hline
\end{tabular}
   }
\end{table}

\begin{table}
\caption{The relative probability densities of the
ten first mass eigenvalues $m_{i,0}$ (in TeV)  for given $x=s/(2\pi R)$
in the tower corresponding to the Case A1
with $kR=6.3$, $\kbar=1$, $\abar_1=0.7$,
$\abar_2=1.3$, $\abar_4=6.6286$.
The highest probability among these ten masses
is normalized to exactly 1 and labeled as such.
Note that the physical masses 
depend on $s_0$, hence on $x$, and are given in terms of the mass eigenvalues by
$m_{i,s_0}=m_{i,0}e^{2\pi x kR}$ in agreement with Eq.\re{masskR2}.
{\label{tableA1prob1}}
}
\vspace{1 cm}
\hspace{-0.5 cm}
\scriptsize
{
\begin{tabular}{|c|c|c|c|c|c|c|c|c|c|c|}
\hline
Mass ${\phantom{[^[_[}}$   & $m_{1,0}$&$m_{2,0}$&$m_{3,0}$&$m_{4,0}$&$m_{5,0}
  $&$m_{6,0}$&$m_{7,0}$&$m_{8,0}$ &$m_{9,0}$ & $m_{10,0}$
\\ \cline{2-11}
 eigenvalues${\phantom{[^[_[}}$  & 0&0.5013&0.766&1.023&1.275&1.525&1.774&2.023&2.271&2.519
\\ \hline\hline
$x$  &\multicolumn{10}{c|}{Relative Probabilities}
\\ \hline
0&1&0.27&0.19&0.14&0.11&0.096&0.083&0.073&0.065&0.059
 \\ \hline
0.4&1&0.27&0.19&0.14&0.11&0.096&0.083&0.073&0.065&0.059
 \\ \hline
0.45&1&0.27&0.19&0.14&0.12&0.084&0.10&0.055&0.090&0.034
 \\ \hline
0.46&1&0.26&0.21&0.11&0.17&0.044&0.18&0.0085&0.23&0.0011
 \\ \hline
0.466&1&0.24&0.25&0.067&0.27&0.0024&0.41&0.035&0.69&0.22
 \\ \hline
0.467&1&0.24&0.26&0.057&0.30&0.000021&0.49&0.071&0.86&0.35
 \\ \hline
0.4676&1&0.24&0.27&0.051&0.33&0.00066&0.55&0.10&0.99&0.45
 \\ \hline
0.4676334&1&0.24&0.27&0.050&0.33&0.00075&0.55&0.10&1.0&0.46
 \\ \hline
0.4676335&1.0&0.24&0.27&0.050&0.33&0.00075&0.55&0.10&1&0.46
 \\ \hline
0.4677&0.98&0.23&0.26&0.049&0.33&0.00095&0.55&0.11&1&0.46
 \\ \hline
0.468&0.92&0.21&0.25&0.042&0.32&0.0021&0.54&0.12&1&0.49
 \\ \hline
0.47&0.56&0.12&0.17&0.013&0.26&0.018&0.50&0.19&1&0.67
 \\ \hline
0.475&0.14&0.022&0.068&0.0017&0.16&0.081&0.42&0.36&0.96&1
 \\ \hline
0.48&0.026&0.0017&0.027&0.012&0.10&0.11&0.31&0.39&0.78&1
 \\ \hline
0.5&0.00012&0.0013&0.0092&0.027&0.067&0.13&0.25&0.41&0.66&1
 \\ \hline
0.9&0.000022&0.0017&0.0092&0.029&0.069&0.14&0.25&0.42&0.67&1
 \\ \hline
0.92&0.000028&0.0022&0.011&0.035&0.083&0.16&0.29&0.46&0.70&1
 \\ \hline
0.94&0.00011&0.0081&0.040&0.11&0.23&0.40&0.59&0.78&0.93&1
 \\ \hline
0.95&0.00052&0.036&0.16&0.40&0.69&0.93&1&0.86&0.56&0.24
 \\ \hline
0.97&0.012&0.51&1&0.58&0.017&0.21&0.40&0.095&0.044&0.26
 \\ \hline
0.99&0.71&0.85&1&0.084&0.24&0.49&0.065&0.13&0.33&0.060
 \\ \hline
0.995&1&0.0075&0.081&0.25&0.31&0.21&0.064&0.000017&0.048&0.13
 \\ \hline
0.999&1&0.26&0.16&0.11&0.073&0.049&0.031&0.019&0.010&0.0044
 \\ \hline
1&1&0.27&0.19&0.14&0.11&0.096&0.083&0.073&0.065&0.059
 \\ \hline
\end{tabular}
   }
\end{table}

\begin{table}
\caption{The relative probability densities of the
ten first physical masses $m_i$ (in TeV)  for given $x_0=s_0/(2\pi R)$
in the tower corresponding to the Case A1,
$\kbar=1$, $\abar_1=0.7$,
$\abar_2=1.3$, $\abar_4=6.6286$.
The highest probability among these ten masses
is normalized to 1 and labeled as such.
When $x_0$ is increased, $\Rbar$ is adjusted according to
\re{masskR2}, keeping $\kbar=1$, so as to
retain the physical masses unchanged.
{\label{tableA1prob2}}
}
\vspace{1 cm}
\hspace{-0.5 cm}
\tiny
{
\begin{tabular}{|c|c|c|c|c|c|c|c|c|c|c|c|}
\hline
 \multicolumn{2}{|c|}{Physical ${\phantom{[^[_[}}$} 
  &  $m_{1}$&$m_{2}$&$m_{3}$&$m_{4}$&$m_{5}$
  &$m_{6}$&$m_{7}$&$m_{8}$ &$m_{9}$ & $m_{10}$
\\ \cline{3-12}
 \multicolumn{2}{|c|}{masses ${\phantom{[^[_[}}$} 
 &      0&0.5013&0.766&1.023&1.275&1.525&1.774&2.023&2.271&2.519
\\ \hline\hline
{\phantom{$\biggl.\biggr.$}}$x_0$ &$\Rbar$&\multicolumn{10}{c|}{Relative Probabilities}
\\ \hline
0       &   6.3 
  &1&0.28&0.19&0.14&0.11&0.096&0.083&0.073&0.065&0.059
    \\ \hline
0.44   & 11.3
  &1&0.27&0.18&0.14&0.11&0.095&0.082&0.072&0.064&0.058
    \\ \hline        
0.475  & 12.0
  &1&0.27&0.19&0.13&0.13&0.077&0.11&0.046&0.1&0.025
    \\ \hline
0.4793 & 12.1
  &1&0.26&0.21&0.11&0.17&0.041&0.18&0.0068&0.24&0.0022
    \\ \hline  
0.4815&12.15
  &1&0.25&0.23&0.081&0.22&0.012&0.3&0.0062&0.47&0.088 
    \\ \hline
0.4828&12.18
  &1&0.24&0.25&0.059&0.29&0.00041&0.45&0.056&0.79&0.29
    \\ \hline
0.4832&12.19
  &1&0.23&0.26&0.051&0.32&0.00033&0.52&0.093&0.94&0.42
    \\ \hline  
0.4834& 12.195 
  &0.97&0.22&0.26&0.045&0.32&0.0015&0.55&0.11&1&0.48 
    \\ \hline  
0.4836& 12.2
  &0.88&0.2&0.24&0.037&0.31&0.0033&0.54&0.13 &1&0.51
    \\ \hline
0.4857& 12.25
  &0.32&0.062&0.11&0.0012&0.21&0.046&0.47&0.28& 1&0.85
    \\ \hline
0.49& 12.3
  &0.095&0.013&0.053&0.0041&0.14&0.09&0.38&0.37&0.9 &1
    \\ \hline
0.61&16.3
  &0.000021&0.0017&0.0088&0.028&0.066&0.14&0.25&0.42&0.66 &1
    \\ \hline  
0.76 & 26.3 
  &0.000022&0.0017&0.0089&0.028&0.067&0.14&0.25&0.42&0.69&1  
    \\ \hline  
0.89  & 56.3 
  &0.000021&0.0016&0.0081&0.025&0.07&0.12&0.25&0.4&0.6&1    
    \\ \hline

\end{tabular}
   }
\end{table}

\newpage

\begin{figure}[ht]
\caption{Case A1 : The logarithm of the field probability
density as a function of
$x=s/(2\pi R)$ for $\kbar=1$, $kR=6.3$, $\abar_1=0.7$,
$\abar_2=1.3$, $\abar_4=6.6286$
and $m_1=0$    \label{figA163m0}
       }

\begin{center}
\epsfxsize=12cm
\epsffile{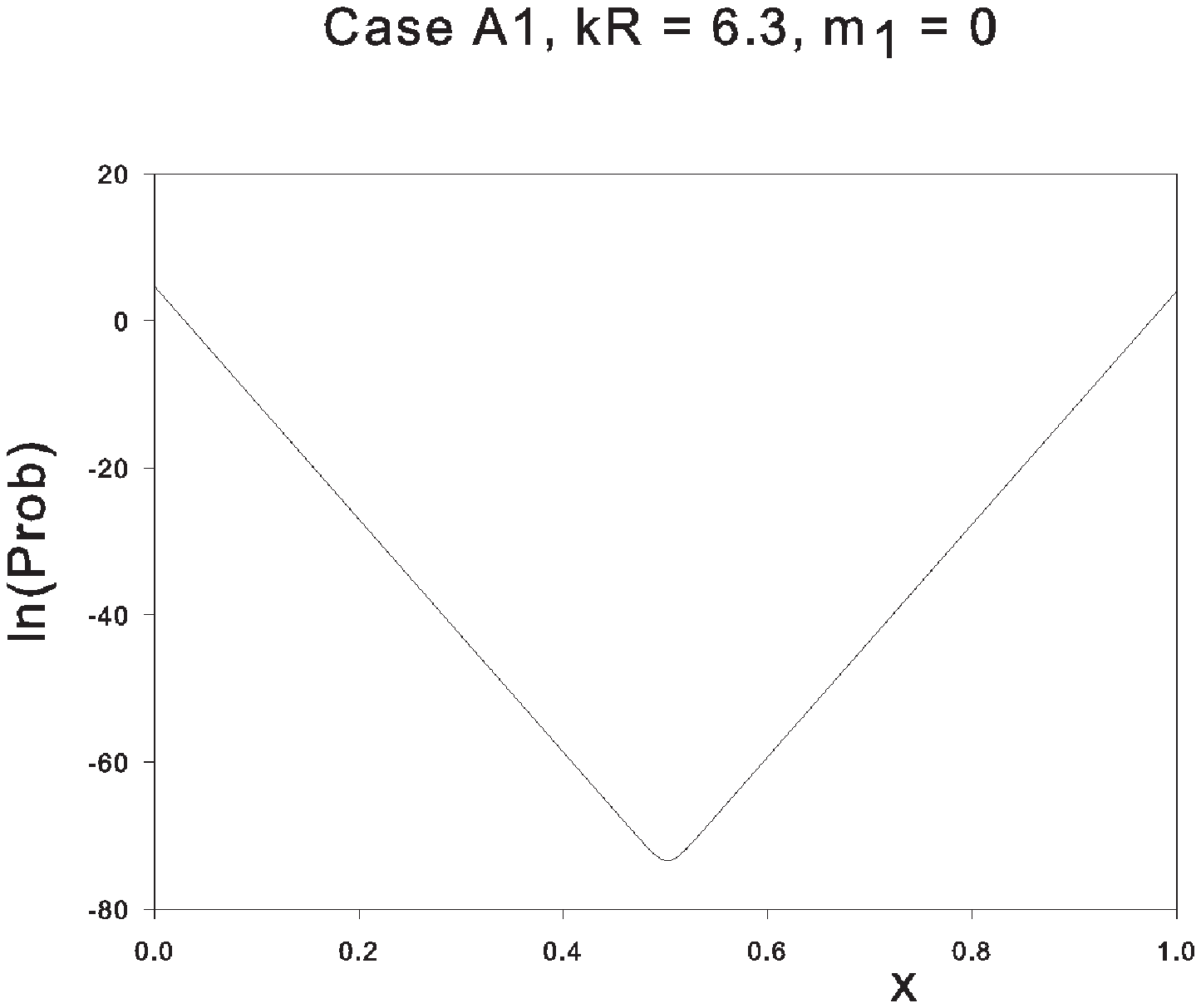}
\end{center}
\end{figure}

\newpage

\begin{figure}[ht]
\caption{Case A1 : The logarithm of the field probability
density as a function of
$x=s/(2\pi R)$ for $\kbar=1$, $kR=6.3$, $\abar_1=0.7$,
$\abar_2=1.3$, $\abar_4=6.6286$
and $m_2=0.501$TeV \label{figA163m501}
       }
\begin{center}
\epsfxsize=12cm
\epsffile{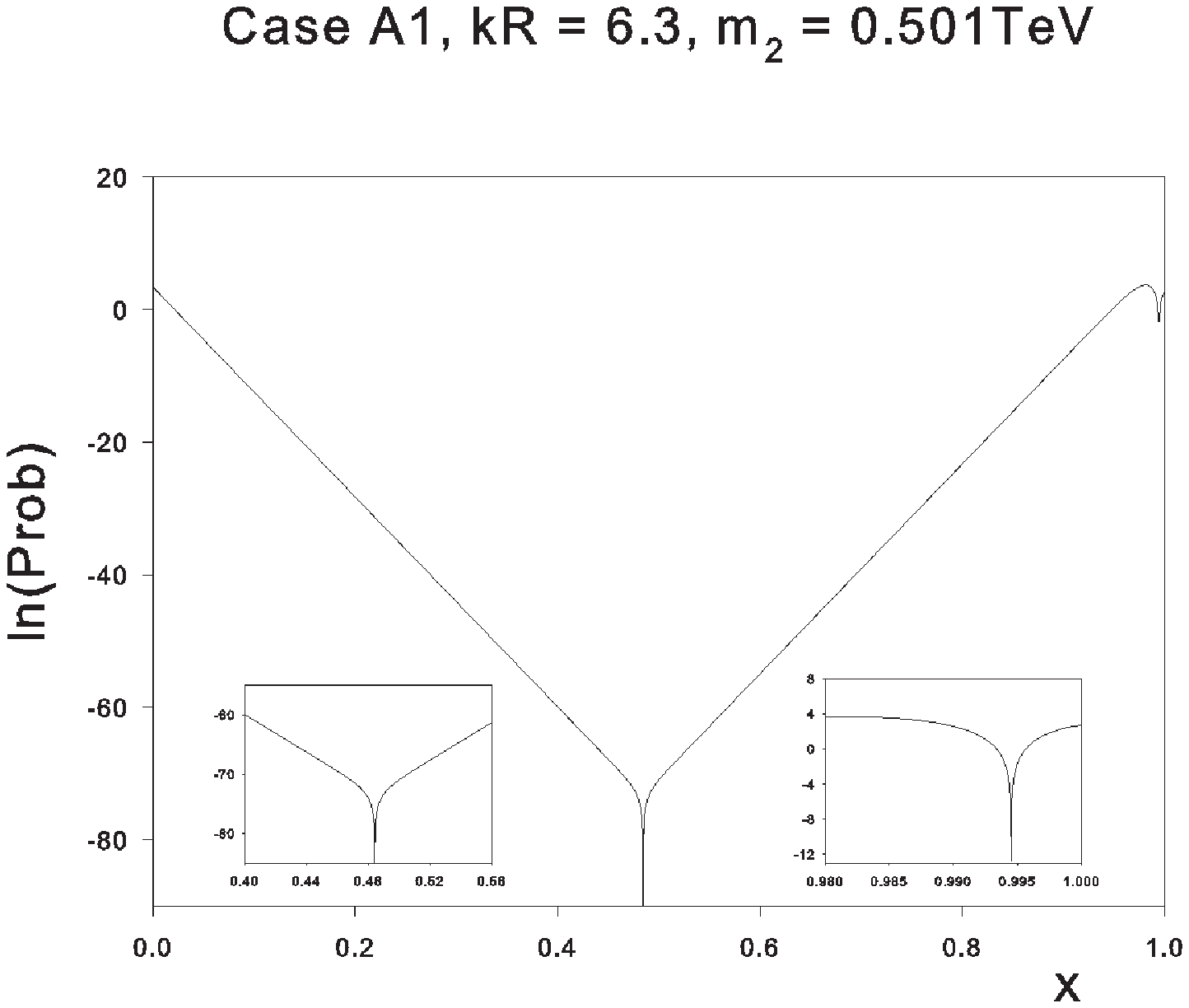}
\end{center}
\end{figure}

\newpage

\begin{figure}[ht]
\caption{Case A1 : The logarithm of the field probability
density as a function of
$x=s/(2\pi R)$ for $\kbar=1$, $kR=6.3$,
 $\abar_1=0.7$,
$\abar_2=1.3$, $\abar_4=6.6286$
and $m_5=1.275$TeV \label{figA163m1275}
       }
\vskip 2 cm
\begin{center}
\epsfxsize=12cm
\epsffile{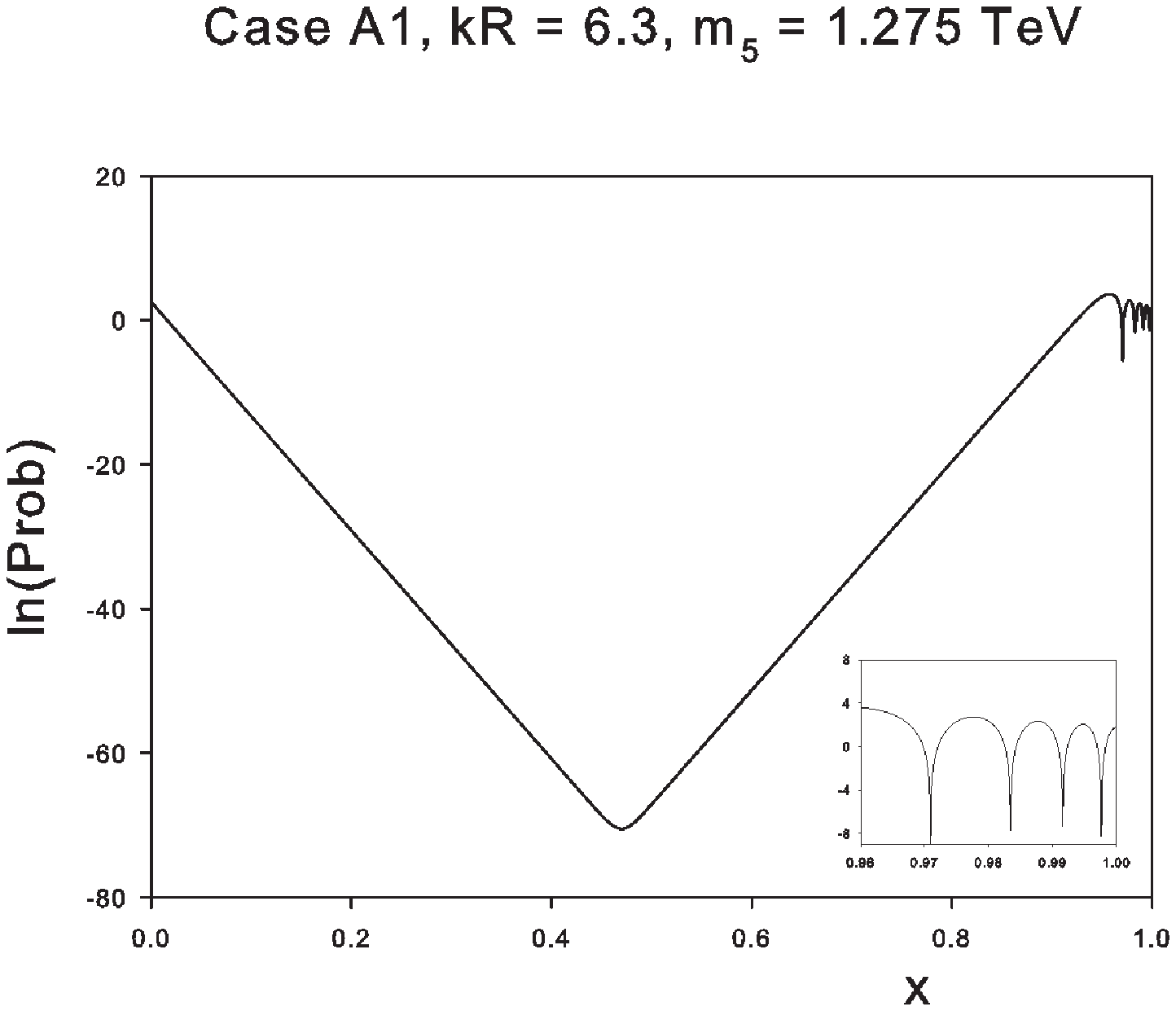}
\end{center}
\end{figure}

\newpage

\begin{figure}[ht]
\caption{Case A4 : The logarithm of the field probability
density as a function of
$x=s/(2\pi R)$ for $\kbar=1$, $kR=6.3$, $\kabar=4$
and $m_1=0$ \label{figA463m0}
       }
\vskip 2 cm
\begin{center}
\epsfxsize=12cm
\epsffile{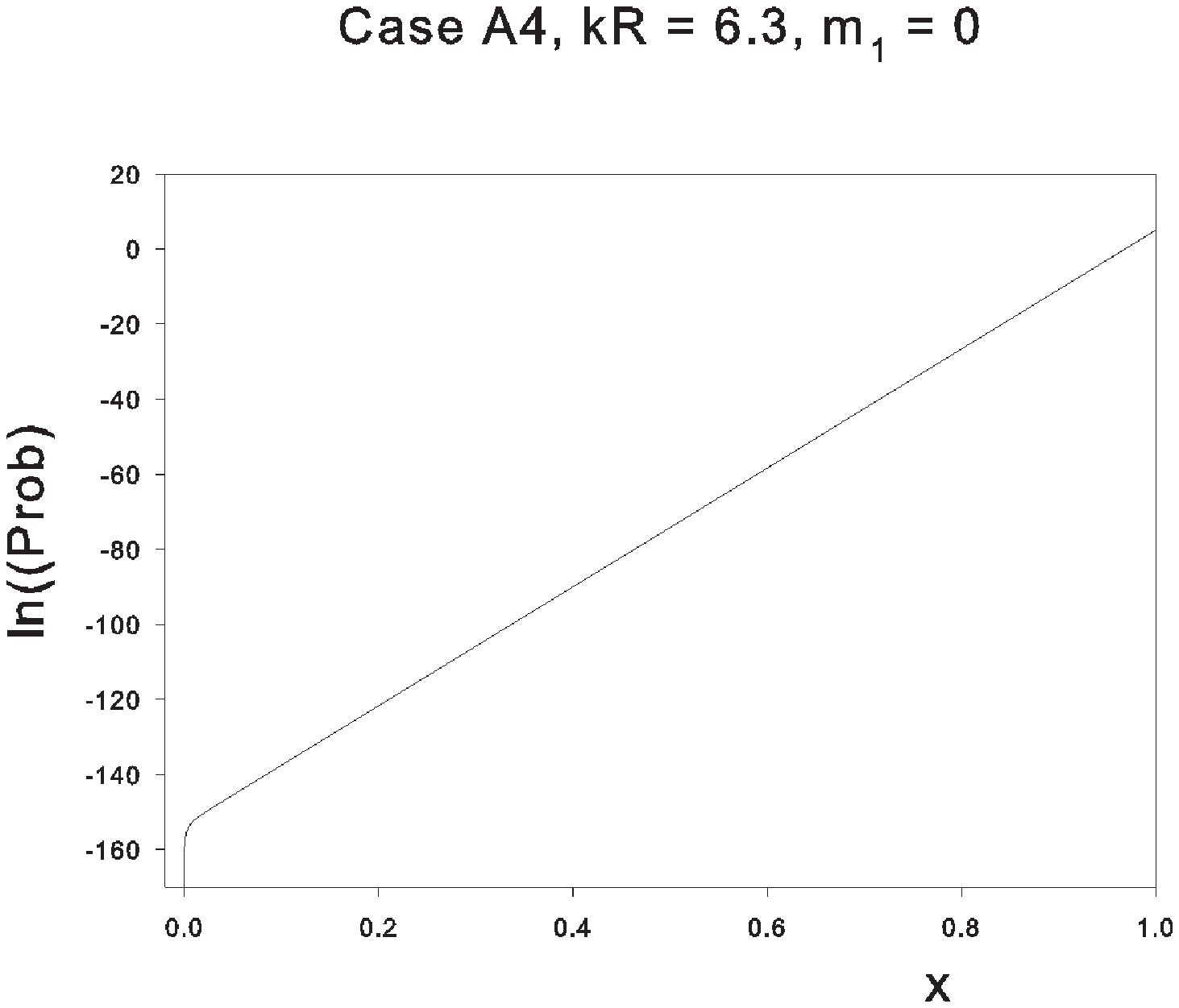}
\end{center}
\end{figure}

\newpage

\begin{figure}[ht]
\caption{Case A4 : The logarithm of the filed probability
density as a function of
$x=s/(2\pi R)$ for $\kbar=1$, $kR=6.3$, $\kabar=4$
and $m_3=0.766$\,TeV \label{figA463m766}
       }
\vskip 2 cm
\begin{center}
\epsfxsize=12cm
\epsffile{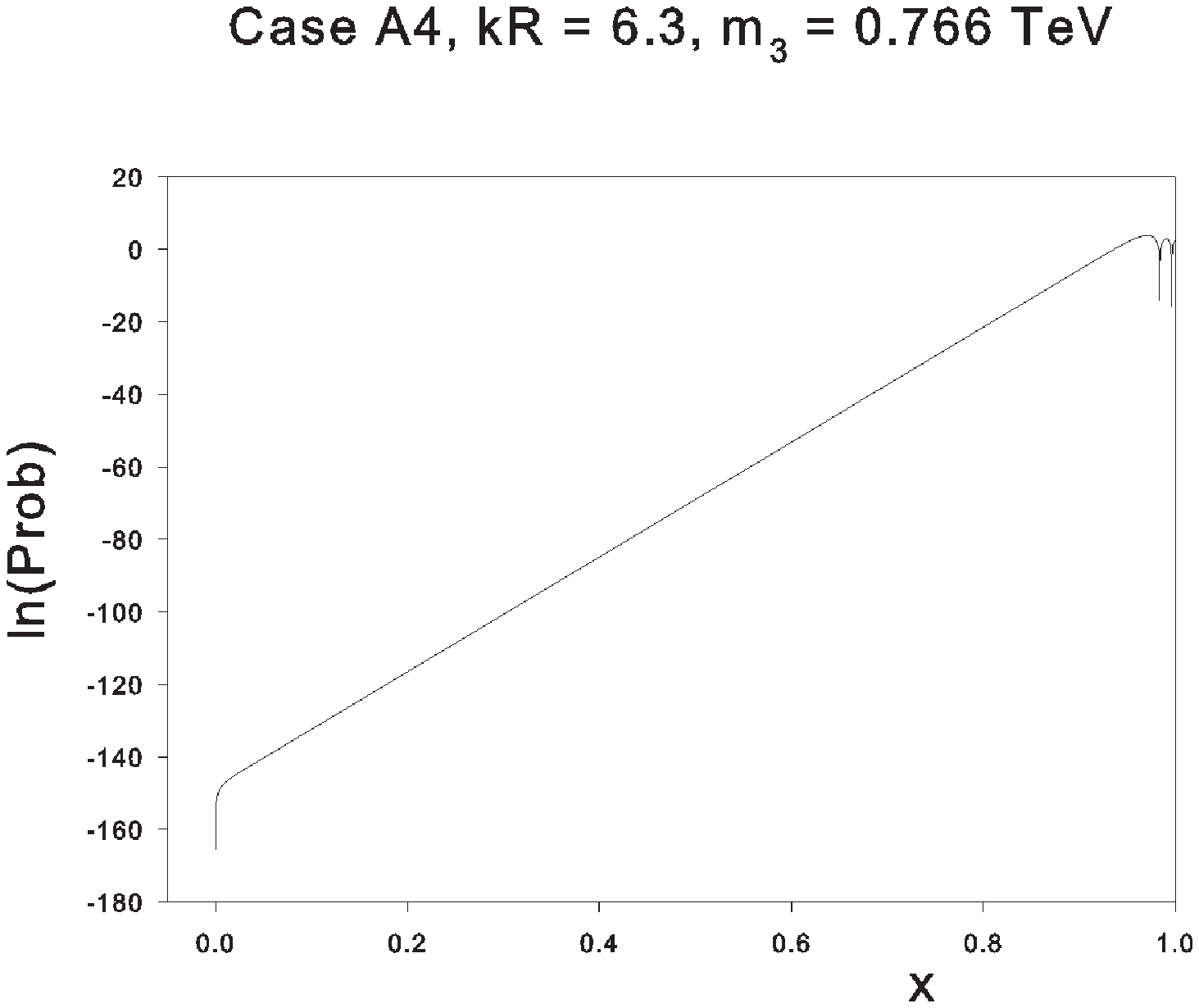}
\end{center}
\end{figure}


\newpage

\begin{thebibliography}{99}

\bibitem{GN}
Grard, F., Nuyts, J.,
{\it{Phys.Rev.}} D 74 , 124013 (2006),
hep-th/0607246

\bibitem{ADD}
Arkhani-Ahmed, N., Dimopoulos, S., Dvali, G.,
{\it{Phys. Lett.}}, {\bf{B429}}, 263 (1998),
hep-ph/9803315, SLAC-PUB-7769, SU-ITP-98/13

Antoniadis, I., Arkani-Hamed, N., Dimopoulos, S., Dvali, G.R.,
{\it{Phys.Lett.}}, {\bf{B436}}, 257 (1998).
hep-ph/9804398,
SLAC-PUB-7801, SU-ITP-98-28, CPTH-S608-0498

\bibitem{KK}
Kaluza, T.,
{\it{Sitzungsber. Preuss. Akad. Wiss. Berlin. (Math. Phys.)}}, 966-972 (1921).
Klein, O.,
{\it{Z. Phys.}} {\bf{37}}, 895-906 (1926).

\bibitem{AGM}
Asorey, M., Garc\'ia \'Alvarez, D., Mu\~{n}oz-Casta\~{n}eda, J.M.,
hep-th/0604089

\bibitem{RS}
Randall, L., Sundrum, R.,
{\it{Phys. Rev. Lett.}} 83 , 3370 (1999)
hep-ph/9905221,
Physical Review Letters 83 , 4690 (1999),
hep-th/9906064

\bibitem{Ben}
Bender, C.M., Boettcher, S.,
{\it{Phys. Rev. Lett.}} 80 , 5243 (1998),
Physics/9712001

\bibitem{FN}
Fairlie, D.B., Nuyts, J.,
{\it{J.Phys.}} {\bf{A38}}, 3611-3624 (2005),
hep-th/0412148


\end{thebibliography}
\end{document}